\documentclass[12pt]{article}

\usepackage{amsthm,amsmath,mathrsfs,amsfonts,amssymb,mathtools,nicefrac}
\usepackage{graphicx,caption,subcaption}
\usepackage{enumerate}
\usepackage{natbib} \RequirePackage[colorlinks,citecolor=blue,urlcolor=blue]{hyperref}
\usepackage{url} 

\usepackage[multiple]{footmisc} \usepackage[capposition=top]{floatrow}

\usepackage{lscape}

\usepackage{xcolor}

\usepackage{adjustbox}
\usepackage{multirow}

\usepackage{amsmath}
\usepackage{accents}
\newlength{\dhatheight}

\newlength{\dcheckheight}

\newcommand{\blind}{1}
\newtheorem{assumption}{Assumption}

\newtheorem{theorem}{Theorem}
\newtheorem{lemma}{Lemma}

\theoremstyle{definition}

\def\N{n}
\def\T{m}

\def\p{p}
\def\q{q}

\begin{document}

\def\spacingset#1{\renewcommand{\baselinestretch}{#1}\small\normalsize} \spacingset{1}

\if1\blind
{
  \title{\bf {\large INFERENCE IN DYNAMIC MODELS FOR PANEL DATA USING THE MOVING BLOCK BOOTSTRAP}}
  \author{ Ayden Higgins\thanks{Address: University of Exeter Business School, Rennes Drive, Exeter EX4 4PU, United Kingdom. E-mail: \texttt{a.higgins@exeter.ac.uk}.} \\ {\small University of Exeter} \and Koen Jochmans\thanks{Address: Toulouse School of Economics, 1 esplanade de l'Universit\'e, 31080 Toulouse, France. E-mail: \texttt{koen.jochmans@tse-fr.eu}.\newline We are grateful to Arthur Lewbel and four anonymous referees, Anders Kock, Roger Moon, Ulrich Mueller, and conference and seminar participants at various institutions for comments and suggestions. \newline Funded by the European Union (ERC-NETWORK-101044319) and by the French Government and the French National Research Agency under the Investissements d' Avenir program (ANR-17-EURE-0010). Views and opinions expressed are however those of the authors only and do not necessarily reflect those of the European Union or the European Research Council. Neither the European Union nor the granting authority can be held responsible for them. } \\  {\small Toulouse School of Economics} 
    }
\date{\small This version: \today} 
  \maketitle
} \fi

\if0\blind
{
  \bigskip
  \bigskip
  \bigskip
  \begin{center}
    {\bf {\normalsize title}}
\end{center}
  \medskip
} \fi
\vspace{-.5cm}
\begin{abstract}
\noindent
Inference in linear panel data models is complicated by the presence of fixed effects when (some of) the regressors are not strictly exogenous. Under asymptotics where the number of cross-sectional observations and time periods grow at the same rate, the within-group estimator is consistent but its limit distribution features a bias term. In this paper we show that a panel version of the moving block bootstrap, where blocks of consecutive cross-sections are resampled with replacement, replicates the limit distribution of the within-group estimator. Confidence ellipsoids and hypothesis tests based on the reverse-percentile bootstrap are thus asymptotically valid without the need to take the presence of bias into account. The bootstrap can equally be used to directly estimate the asymptotic bias and variance.
\end{abstract}




\spacingset{1.45}

\renewcommand{\theequation}{\arabic{section}.\arabic{equation}}  \setcounter{equation}{0}

\newpage

\section{Introduction}
Econometric models for panel data almost invariably feature fixed effects. Their presence complicates estimation and inference as they introduce bias in the fixed-effect estimator of the parameter of interest, in general. This bias remains important in large samples unless the number of time-series observations, $\T$, is large relative to the cross-sectional sample size, $\N$, in the sense that $\nicefrac{\N}{\T}\rightarrow 0$ as $\N,\T\rightarrow\infty$. This is not usually the case. 
Several ways to correct for the bias have been proposed; see \cite{ArellanoHahn2007} for an overview. Doing so recenters the estimator's limit distribution around zero,  thereby restoring the validity of inference procedures based on it.

An alternative approach, which has been explored by \cite{GoncalvesKaffo2015} and \cite{HigginsJochmans2024}, is to use the bootstrap to replicate the distribution of the fixed-effect estimator, including its bias. This, then, allows us to perform inference based on the bootstrap distribution in the usual manner; no adjustment for the presence of bias needs to be made. The difficulty here lies in constructing a bootstrap scheme that correctly reproduces the bias. \cite{HigginsJochmans2024} show that the parametric bootstrap does so in quite general nonlinear models. Outside the parametric framework, however, there are currently results only for the linear autoregressive model. In particular, \cite{GoncalvesKaffo2015} show that a recursive version of the wild bootstrap permits inference using the within-group estimator. They also demonstrate the failure of several other conventional bootstrap schemes to replicate bias in their setting, illustrating the difficulty in devising a successful procedure.

In this paper we study the linear model where the errors are allowed to be correlated with both past and future regressors. This is a workhorse model in applied economics. Indeed, many regressors react to past outcomes, creating feedback. Microeconometric examples are fertility decisions in wage equations, labor-force participation decisions, and household demand systems (see \citealt{ArellanoCarrasco2003} and \citealt{BrowningLechene2003}  for discussions). A macroeconometric example is panel local projections (\citealt{JordaTaylor2025}, \citealt{MeiShengShi2025}). As in the autoregressive problem of \cite{GoncalvesKaffo2015}, which is nested in our setup, the bias in the within-group estimator arises due to the correlation between the time-demeaned regressors and the errors. Its precise form, however, depends on the feedback process. A bootstrap designed around any form of recursion requires this process as an input and will, therefore, not be helpful in our setting. Whether a bootstrap scheme that mimics the bias for feedback processes of an unknown form can be constructed is not known.

We show that a version of the moving block bootstrap of \cite{Kunsch1989}, where blocks of consecutive cross-sections are resampled with replacement, replicates the distribution of the within-group estimator under asymptotics where $\nicefrac{\N}{\T} \rightarrow c \in [0,+\infty)$. The potential for this bootstrap scheme to replicate bias in our context has not previously been explored. Although  \cite{Goncalves2011} has shown that it is capable of yielding correct inference in our model, she did so under rate conditions on $\N$ and $\T$ that ensure that the bias is asymptotically negligible.

Whereas we focus on a setting with cross-sectional independence, \cite{Goncalves2011} provided limit theory under various degrees of cross-sectional dependence. In her setting, the normal equations are assumed to converge at the rate $\nicefrac{1}{\sqrt{\N^\rho\T}}$ for some $\rho\in [0,1]$, and asymptotic bias is assumed away by imposing that ${\nicefrac{\N^\rho}{\T}}\rightarrow 0$.  As $\rho$ decreases and the amount of dependence increases, the convergence rate of the estimator goes down, and the less pertinent the presence of bias at order $\nicefrac{1}{\T}$ becomes. In the most severe case we have $\rho=0$; the cross-sectional sample size becomes irrelevant in the asymptotic analysis and no bias issue arises. 
Because our bootstrap resamples entire cross-sections, it preserves the cross-sectional dependence present in the data. As the bias does not involve any cross-unit moments, its form is not affected by the presence of cross-sectional dependence. Therefore, it continues to correctly replicate the (possibly biased) limit distribution of the within-group estimator.
A similar robustness result holds with respect to the inclusion of time effects. Under cross-sectional independence, such effects do not affect the asymptotic bias (see \citealt{HahnMoon2006} for a derivation in the pure autoregressive case). 
We illustrate the robustness in simulations.

This paper's contribution should be interpreted against the backdrop of a recent interest in the validity of the bootstrap in the presence of (asymptotic) bias; see, notably, \cite{CattaneoJansson2018} and \cite{CavaliereGoncalvesNielsenZanelli2024} for contributions outside the panel data context. 
We conjecture that the validity of our bootstrap scheme extends to a large class of nonlinear problems, at the level of generality of \cite{HahnKuersteiner2011}. We have established this to be the case for the variance estimator in a semiparametric version of the many normal means problem of \cite{NeymanScott1948}, where the distribution of the data is unspecified and observations are stationary over time. This is a well-known test case for nonlinear problems. A general theory is left for future work.

The remainder of the paper is structured as follows. 
In the next section we first formally state our model and assumptions, and derive the limit distribution of the within-group estimator. This yields a somewhat more general expression of its asymptotic bias than is available in the literature. The following section introduces the moving block bootstrap and states our main result---that is, the distribution of the bootstrap within-group estimator, centered around the within-group estimator and conditional on the data, is consistent for the limit distribution of the latter estimator centered around the truth---together with its chief implications for estimation and inference. The penultimate section reports on numerical experiments in the baseline case as well as in situations with time effects and cross-sectional dependence. The final section contains concluding remarks. All proofs are in the Appendix.

\section{Linear regression with fixed effects}
We are interested in estimation of and inference on the slope vector $\beta$ in the linear model
$$
y_{it} = \alpha_i + x_{it}^\prime \beta + \varepsilon_{it}, \qquad \mathbb{E}(\varepsilon_{it}) = 0, \qquad \mathbb{E}(x_{it} \varepsilon_{it}) = 0,
$$
from $\N\times \T$ panel data, treating the $n$ intercepts $\alpha_1,\ldots, \alpha_n$ as unknown parameters to be estimated.

A key feature of our model is that the regressors $x_{it}$ are not assumed to be strictly exogenous. This is important. For example, consider the simple first-order autoregression
$$
y_{it} = \alpha_i + y_{it-1} \beta + \varepsilon_{it}
$$
with uncorrelated errors. This is a simple feedback process, as $x_{it} = y_{it-1}$, that has received particular attention. When errors are homoskedastic with variance $\sigma^2$, it is easy to see that 
$\mathbb{E}(x_{it} \varepsilon_{it+\tau})=0$ but $\mathbb{E}(x_{it+\tau} \varepsilon_{it}) = \beta^{\tau-1}\sigma^2\neq 0$ for all $\tau>0$. A more general feedback process is obtained when $y_{it}$ and $x_{it}$ follow a vector autoregressive process. Of course, feedback need not be restricted to be linear. Indeed, $x_{it}$ could contain choice variables, such as fertility or schooling decisions, that are affected by past outcomes, such as wages. A main driver behind this paper is the desire to remain agnostic about the specific form of the feedback process when performing inference.


To establish our theoretical results we will work under three conventional assumptions. The first of these contains moment requirements and mixing conditions on the time-series processes. 

\begin{assumption} \label{ass:ass1}
For some $r>2$, 

\medskip\noindent
(i)~The variables $x_{it}$ and $\varepsilon_{it}$ have uniformly bounded moments of order $2r$. 

\medskip\noindent
(ii)~The variables $x_{it} \varepsilon_{it}$ have uniformly bounded moments of order $2r$. 

\medskip\noindent
(iii)~The processes $\lbrace (x_{it}, \varepsilon_{it}) \rbrace_t$ are independent across $i$.



\medskip\noindent
(iv)~For each $i$, $\lbrace (x_{it},\varepsilon_{it}) \rbrace_t$ is stationary and strong mixing with the mixing coefficients, $a_i$, satisfying
$$
\sup_{1\leq i \leq \N} a_i(h) = O(h^{-s})
$$
\hphantom{(v)~}for some $s > \nicefrac{4r}{r-2}$.\footnote{Here,
$$
a_i(h) \coloneqq
\sup_{1\leq t \leq \T} 
\sup_{A\in\mathcal{A}_{it}}
\sup_{B\in \mathcal{B}_{it+h}}
\lvert
\mathbb{P}(A\cap B)
-
\mathbb{P}(A) \, \mathbb{P}(B)  
\rvert,
$$
for $\mathcal{A}_{it}$ and $\mathcal{B}_{it}$ the sigma algebras generated by the sequences $(x_{it},\varepsilon_{it}) ,(x_{it-1},\varepsilon_{it-1}),\ldots$ and $(x_{it},\varepsilon_{it}) ,(x_{it+1},\varepsilon_{it+1}),\ldots$, respectively.}

\end{assumption}


\noindent
Conditions of the type stated in Assumption \ref{ass:ass1} are frequent when analyzing least squares estimators from dependent data; see, e.g., \cite{White1984}. The strong-mixing requirements are well-known to hold for a variety of processes, including autoregressive/moving-average processes  \citep{Mokkadem1988}. The assumption rules out time effects and cross-sectional dependence. This is largely done for simplicity. We will investigate the robustness of our results below.

The second assumption states standard rank conditions on two covariance matrices. Here and in the sequel we let $z_{it} \coloneqq x_{it} - \mathbb{E}(x_{it})$.

\begin{assumption} \label{ass:ass2}
\mbox{}
As $\N,\T\rightarrow \infty$, 
$$
\varSigma_{\N,\T} \coloneqq \frac{1}{\N\T} \sum_{i=1}^{\N} \sum_{t=1}^{\T} 
\mathbb{E}
(
z_{it}^{\vphantom{\prime}} z_{it}^\prime
)
$$
and 
$$
\varOmega_{\N,\T}
\coloneqq
\frac{1}{\N\T} \sum_{i=1}^{\N} \sum_{t=1}^{\T}
\left(
\mathbb{E}(z_{it}^{\vphantom{\prime}} z_{it}^\prime  \varepsilon_{it}^{2})
+
\sum_{\tau=1}^{\T-1} 
\left(1 - \nicefrac{\tau}{\T} \right)
\
\mathbb{E}
(
(z_{it}^{\vphantom{\prime}} z_{it+\tau}^\prime + z_{it+\tau}^{\vphantom{\prime}} z_{it}^\prime) \,
\varepsilon_{it}^{\vphantom{\prime}} \varepsilon_{it+\tau}^{\vphantom{\prime}} 
)
\right)
$$
converge to finite positive-definite matrices.

\end{assumption}

The third assumption states the asymptotic approximation under which we proceed.

\begin{assumption} \label{ass:ass3}
$\N,\T\rightarrow\infty$ with $\nicefrac{\N}{\T} \rightarrow c \in [0,+\infty)$.
\end{assumption}

\noindent
Because we allow for $\mathbb{E}(x_{it} \varepsilon_{it+\tau})$ and $\mathbb{E}(x_{it+\tau} \varepsilon_{it})$ to be non-zero when $\tau\neq 0$, both the within-group least-squares estimator and generalized method-of-moment estimators as in \cite{ArellanoBond1991} will be inconsistent, in general, under asymptotics where $\T$ is held fixed. Assumption \ref{ass:ass3} thus demands $\T$ to grow along with $\N$. It nonetheless permits $\N$ and $\T$ to grow at the same rate, and so our focus is on panels where $\N$ is not small relative to $\T$.


The within-group estimator of $\beta$ is 
$$
\hat{\beta}
\coloneqq
\left(\sum_{i=1}^{\N} \sum_{t=1}^{\T} (x_{it} - \bar{x}_i ) (x_{it} - \bar{x}_i )^\prime  \right)^{-1}
\left(\sum_{i=1}^{\N} \sum_{t=1}^{\T} (x_{it} - \bar{x}_i ) (y_{it} - \bar{y}_i )^{\hphantom{\prime}} \right),
$$
where $\bar{y}_i \coloneqq \nicefrac{1}{m} \sum_{t=1}^m y_{it}$ and $\bar{x}_i \coloneqq \nicefrac{1}{m} \sum_{t=1}^m x_{it}$. Under Assumptions \ref{ass:ass1}-\ref{ass:ass3} this estimator is consistent and asymptotically normally-distributed, but its limit distribution features an asymptotic bias term unless $\nicefrac{\N}{\T}\rightarrow 0$. The bias arises from the fact that the normal equations factor as
$$
\sum_{i=1}^{\N} \sum_{t=1}^{\T} (x_{it}-\bar{x}_i) \varepsilon_{it}
=
\sum_{i=1}^{\N} \sum_{t=1}^{\T} z_{it} \varepsilon_{it}
\\
 -
\sum_{i=1}^{\N} \T   \mathrm{cov}(\bar{z}_i,\bar{\varepsilon}_i) 
+ o_P(\sqrt{\N\T}),
$$
where we let $\bar{z}_i \coloneqq \nicefrac{1}{\T} \sum_{t=1}^{\T} z_{it}$ and $\bar{\varepsilon}_i \coloneqq \nicefrac{1}{\T} \sum_{t=1}^{\T} \varepsilon_{it}$. 
While the first right-hand side term in this expression is mean zero and, after scaling, satisfies a central limit theorem, the second right-hand side term will shift the limit distribution away from zero, thereby causing bias.


\begin{theorem} \label{thm:thm1}
Let Assumptions \ref{ass:ass1}-\ref{ass:ass3} hold. Then
$$
\sqrt{\N\T} \, 
(\hat{\beta} - \beta 
)
-
\sqrt{\frac{\N}{\T}} \varSigma^{-1}_{\N,\T} b_{\N,\T}^{\vphantom{-1}}
\overset{L}{\rightarrow} N(0,\varUpsilon),
$$
where
$
b_{\N,\T}
\coloneqq
-
\nicefrac{1}{\N} \sum_{i=1}^n
\sum_{\tau=1}^{m-1} \nicefrac{(\T-\tau)}{\T}
\left(
\mathbb{E}(z_{it}\varepsilon_{it+\tau}) + \mathbb{E}(z_{it+\tau}\varepsilon_{it}) 
\right)
$
and
$
\varUpsilon \coloneqq \mathrm{lim}_{\N,\T\rightarrow\infty} \varUpsilon_{\N,\T}
$
for 
$
\varUpsilon_{\N,\T} \coloneqq \varSigma_{\N,\T}^{-1} \varOmega_{\N,\T}^{\vphantom{-1}} \varSigma_{\N,\T}^{-1}.
$
\end{theorem}

\noindent
Theorem \ref{thm:thm1} generalizes results for models with predetermined regressors generated through a specified linear process such as those derived in \cite{Nickell1981}, \cite{HahnKuersteiner2002} and \cite{AlvarezArellano2003} for first-order autoregressive processes,  \cite{DhaeneJochmans2016b} for vector autoregressive processes, and \cite{ChudikPesaranYang2018} for (possibly infinite) moving-average processes.

In light of Theorem \ref{thm:thm1}, one way forward would be to estimate the bias term $b_{\N,\T}$ by means of a conventional HAC estimator, using plug-in estimators of the covariances appearing in it based on the fixed-effect estimator of the parameters $\alpha_1,\ldots, \alpha_{\N}$ and $\beta$, and then recenter the within-group normal equations around it.  \cite{HahnKuersteiner2011} proposed using
$$
\hat{b}_{\N,\T}
\coloneqq - \frac{1}{\N\T} \sum_{i=1}^\N \sum_{\tau=-h}^h 
\left(
  \sum_{t=\text{max}\lbrace 1, 1+\tau \rbrace}^{\text{min}\lbrace \T, \T+\tau \rbrace} (x_{it-\tau} - \bar{x}_i) \, \hat{\varepsilon}_{it}
\right),
$$
where $\hat{\varepsilon}_{it} \coloneqq (y_{it} - \bar{y}_i) - (x_{it}-\bar{x}_i)^\prime \hat{\beta}$ are the within-group residuals and $h$ is a chosen bandwidth parameter that grows with $\T$ so that $\nicefrac{h}{\T}\rightarrow 0$ (\citealt[Theorem 2]{HahnKuersteiner2011}). Under regularity conditions this yields an estimator that is asymptotically unbiased. An alternative would be to use a jackknife estimator as in \cite{DhaeneJochmans2015}. In either case, asymptotically-valid inference can then be performed by relying on the limit distribution in the theorem, complemented with a suitable HAC estimator of the asymptotic variance that features in it. 
As an alternative, in the next section we demonstrate that a bootstrap procedure applied to the within-group estimator can be used to perform inference directly, without the need to estimate either the bias or the variance.

\section{Moving block bootstrap}
Our bootstrap scheme consists of applying the moving block bootstrap of \cite{Kunsch1989} in the time-series dimension of the data, jointly for all cross-sectional units. For integers $p$ and $q$ with $\T = \p\times \q$, we randomly select $\p$ blocks of $\q$ consecutive cross-sections from the original data; the blocks may overlap. Our bootstrap sample is then obtained on concatenating these $\p$ blocks. If we let $\varpi_1,\ldots , \varpi_{\p}$ be a random sample from the discrete uniform distribution on $\lbrace 0,\ldots, \T-\q \rbrace$, the bootstrap time series $\lbrace (y_{it}^*, x_{it}^*) \rbrace$ is generated as
$$
y_{i \, (\p^\prime-1)\q+\q^\prime}^* \coloneqq y_{i\, \varpi_{\p^\prime}+\q^\prime}, 
\qquad
x_{i \, (\p^\prime-1)\q+\q^\prime}^* \coloneqq x_{i\, \varpi_{\p^\prime}+\q^\prime},
$$
for $1\leq \p^\prime \leq \p$ and $1\leq \q^\prime \leq \q$. 
Here, the random variables $\varpi_1,\ldots, \varpi_p$ select starting points for the different blocks. When the block length, $q$, is set to one this reduces to \citeauthor{Efron1979}'s (\citeyear{Efron1979}) bootstrap applied to the cross-sections. Given the presence of time-series dependence, however, we will require $\q$ to grow with $\T$ for the bootstrap to work in our context.

The within-group estimator computed from the sample generated via our bootstrap is
$$
\hat{\beta}^*
\coloneqq
\left(\sum_{i=1}^{\N} \sum_{t=1}^{\T} (x_{it}^* - \bar{x}_i^* ) (x_{it}^* - \bar{x}_i^* )^\prime  \right)^{-1}
\left(\sum_{i=1}^{\N} \sum_{t=1}^{\T} (x_{it}^* - \bar{x}_i^* ) (y_{it}^* - \bar{y}_i^* )^{\hphantom{\prime}} \right),
$$
where $\bar{y}_i^* \coloneqq \nicefrac{1}{m} \sum_{t=1}^m y_{it}^*$ and $\bar{x}_i^* \coloneqq \nicefrac{1}{m} \sum_{t=1}^m x_{it}^*$.

The following theorem is our main result. In it, as usual, we let $\mathbb{P}^*$ denote a probability computed with respect to the bootstrap measure, that is, conditional on the original data.

\begin{theorem} \label{thm:thm2}
Let Assumptions \ref{ass:ass1}-\ref{ass:ass3} hold and suppose that $\q\rightarrow \infty$ with {$q = o({\T})$}. Then
$$
\mathbb{P}
\left(
\sup_a
\left\lvert
\mathbb{P}^*(\sqrt{\N\T}(\hat{\beta}^*-\hat{\beta})\leq a)
-
\mathbb{P}(\sqrt{\N\T}(\hat{\beta}-\beta)\leq a)
\right\rvert
> \epsilon
\right)
=
o(1)
$$	
for any $\epsilon>0$.
\end{theorem}

\noindent
This theorem extends the findings in \cite{Goncalves2011}, where Assumption \ref{ass:ass3} is replaced by the stronger requirement that  $\nicefrac{\N}{\T}\rightarrow 0$. Because the bias in $\sqrt{\N\T}(\hat{\beta}-\beta)$ is of the order $\sqrt{\nicefrac{\N}{\T}}$, this rate condition ensures that the limit distribution of the within-group estimator does not feature a bias term.  


The intuition behind Theorem \ref{thm:thm2} is as follows. 
The bootstrap normal equations  satisfy
\begin{equation*}
\begin{split}
\sum_{i=1}^{\N} \sum_{t=1}^{\T} (x_{it}^*-\bar{x}_i^*) \hat{\varepsilon}_{it}^*
& =
\sum_{i=1}^{\N} \sum_{t=1}^{\T} (z_{it}^* \varepsilon_{it}^* - z_{it} \varepsilon_{it})
 -
\sum_{i=1}^{\N} 
\T \mathrm{cov}^*(\bar{z}_i^*,\bar{\varepsilon}_i^*) 
+ o_{P^*}(\sqrt{\N\T}),
\end{split}
\end{equation*}
where
$
\hat{\varepsilon}_{i \, (\p^\prime-1)\q+\q^\prime}^* \coloneqq \hat{\varepsilon}_{i\, \varpi_{\p^\prime}+\q^\prime}, 
$
for $1\leq \p^\prime \leq \p$ and $1\leq \q^\prime \leq \q$ and we let $\bar{z}_i^* \coloneqq \nicefrac{1}{m} \sum_{t=1}^m z_{it}^*$ and $\bar{\varepsilon}_i^* \coloneqq \nicefrac{1}{m} \sum_{t=1}^m \varepsilon_{it}^*$. The $ \T \mathrm{cov}^*(\bar{z}_i^*,\bar{\varepsilon}_i^*)$ are unit-specific moving-block bootstrap estimators of the covariances $ \T \mathrm{cov}(\bar{z}_i,\bar{\varepsilon}_i)$  (see, e.g., \citealt{GoncalvesWhite2002}) which cause the bias in Theorem \ref{thm:thm1}. Indeed, after scaling, the first term on the right-hand side yields the limit distribution that features in Theorem \ref{thm:thm1}, whereas the second right-hand side term is a consistent estimator of the bias term $b_{\N,\T}$, ensuring that the bootstrap distribution is correctly centered.


Theorem \ref{thm:thm2} has a number of useful implications. The first, and most striking, is that inference based on the reverse-percentile bootstrap is valid in spite of the presence of asymptotic bias. To see this,
suppose we wish to perform inference on linear contrasts of the form $\theta\coloneqq \Delta^\prime \beta$, where $\Delta$ is a chosen vector of conformable dimension. For $\alpha \in (0,1)$, let
$$
\hat{Q}_\alpha \coloneqq \inf \lbrace Q: \alpha \leq \mathbb{P}^*(\hat{\theta}^*-\hat{\theta} \leq Q)   \rbrace,
$$
with $\hat{\theta}\coloneqq \Delta^\prime \hat{\beta}$  and $\hat{\theta}^*\coloneqq \Delta^\prime \hat{\beta}^*$. By \citet*[Lemma 23.3]{vanderVaart2000} we have that, under the conditions of Theorem \ref{thm:thm2},
$$
\lim_{\N,\T\rightarrow\infty}
\mathbb{P}
(
\hat{\theta} - \theta \leq \hat{Q}_\alpha
)
=
\alpha ,
$$
which allows the construction of confidence intervals and decision rules to conduct inference on $\theta$. For example, a one-sided confidence interval with nominal coverage $\alpha$ is the interval
$$
\lbrace \vartheta: \hat{\theta} - \hat{Q}_{\alpha} \leq \vartheta  \rbrace .
$$
More general confidence sets and decision rules for hypothesis testing follow in the same way.

Theorem \ref{thm:thm2} permits inference without relying on the normal approximation given in Theorem \ref{thm:thm1}. Moreover, it requires neither an estimator of the bias nor an estimator of the variance of $\hat\beta$. This makes inference particularly straightforward. Nevertheless, if estimators of the bias and variance are desired, the bootstrap can be used to construct them. Such estimators are bootstrap alternatives to the ones already available in the literature. We describe how to obtain them next.


First, Theorems \ref{thm:thm1} and \ref{thm:thm2} imply that the bootstrap median consistently estimates the bias term in Theorem \ref{thm:thm1}. Hence,
$$
\check{\theta} \coloneqq \hat{\theta} - \hat{Q}_{\nicefrac{1}{2}} 
$$
is a bootstrap-based bias-corrected estimator of $\theta$ with the same properties as the corrected estimators discussed above.


If desired, the variance of the limit distribution in Theorem \ref{thm:thm1} may equally be estimated by means of the bootstrap. This is similar to a HAC estimator. To describe it, we introduce
$$
\hat{\varOmega}_{\N,\T}^*
\coloneqq
\frac{1}{\N\p} \sum_{i=1}^{\N} \sum_{\p^\prime =1 }^{\p} V_{i \varpi_{\p^\prime}}^{\vphantom{\prime}} V_{i\varpi_{\p^\prime}}^\prime,
\qquad
V_{i\varpi} \coloneqq 
\left(
\frac{1}{\sqrt{\q}} \sum_{\q^\prime=1}^{\q} (x_{i\varpi+\q^\prime} - \bar{x}_i^*) \,  \hat{\epsilon}_{i\varpi+\q^\prime}^*
\right),
$$
where $\hat{\epsilon}_{it}^* \coloneqq (y_{it}- \bar{y}_i^*) - (x_{it}- \bar{x}_i^*)^\prime\hat{\beta}^*$ are bootstrap residuals.
This is an estimator of the (conditional) bootstrap variance
$$
\hat{\varOmega}_{\N,\T}
\coloneqq
\mathrm{var}^*
\left( 
\frac{1}{\sqrt{\N\T}} \sum_{i=1}^{\N} \sum_{t=1}^{\T} (x_{it}^* - \bar{x}^*_i) \, \hat{\varepsilon}_{it}^*
\right) 
$$
which, in turn, is an estimator of $\varOmega_{\N,\T}$.
If we further
introduce the shorthands
$$
\hat{\varSigma}_{\N,\T}^* \coloneqq 
\frac{1}{\N\T} \sum_{i=1}^{\N} \sum_{t=1}^{\T} (x_{it}^*-\bar{x}_{i}^*) (x_{it}^*-\bar{x}_{i}^*)^\prime
,
\quad
\hat{\varSigma}_{\N,\T} \coloneqq 
\frac{1}{\N\T} \sum_{i=1}^{\N} \sum_{t=1}^{\T} (x_{it}-\bar{x}_{i}) (x_{it}-\bar{x}_{i})^\prime,
$$
we can define
$$
\hat{\varUpsilon}^*_{\N,\T} \coloneqq (\hat{\varSigma}_{\N,\T}^*)^{-1} \hat{\varOmega}_{\N,\T}^{*} (\hat{\varSigma}_{\N,\T}^*)^{-1},
\qquad
\hat{\varUpsilon}_{\N,\T} \coloneqq \hat{\varSigma}_{\N,\T}^{-1} \hat{\varOmega}_{\N,\T}^{\vphantom{-1}} \hat{\varSigma}_{\N,\T}^{-1}.
$$
This is an adaptation of the variance estimator proposed in \cite{Goncalves2011}  to the case of cross-sectionally independent data. By the same arguments as those used in \citet[Theorem B.1 and Lemma B.1.(iii)]{Goncalves2011} it can be verified that
$
\hat{\varUpsilon}^*_{\N,\T} - \varUpsilon \overset{P^*}{\rightarrow} 0 
$
and that
$\hat{\varUpsilon}_{\N,\T} - \varUpsilon \overset{P}{\rightarrow} 0$.\footnote{This result, combined with Theorem \ref{thm:thm2}, can be used to justify the use of the reverse-percentile bootstrap to estimate the quantiles of $\sigma^{-1}\sqrt{\N\T}(\hat\theta - \theta)$, where $\sigma^2 \coloneqq \Delta^\prime \varUpsilon \Delta$. However, we do stress that, due to the presence of bias in the limit distribution of the within-group estimator, the normalized statistic is not properly studentized and, hence, does not follow a pivotal limit distribution. Consequently, the usual argument for obtaining asymptotic refinements via this route does not apply in the context under study here.}

\section{Simulations}
In this section we provide simulation results for four different situations. All four are built around the autoregressive problem 
$$
y_{it} = y_{it-1} \beta + \varepsilon_{it},
$$
with $\beta = \nicefrac{1}{2}$.\footnote{An earlier version of this paper contained simulation results for other forms of feedback; a binary regressor depending on the previous outcome and a regressor generated jointly with the outcome through a vector autoregression.} The within-group estimator is invariant to the distribution of the fixed effects, so setting $\alpha_i = 0$ for all $1\leq i \leq \N$ is without loss of generality. The first situation serves to validate our baseline setting. Here we draw $\varepsilon_{it}\sim\mathrm{i.i.d.}~N(0,1)$ and then estimate $\beta$ by the within-group estimator. The second situation uses the same data generating process but controls for aggregate time effects through the inclusion of time dummies in the estimator. The third and fourth situations consider the same estimators as the first and second, respectively, but feature cross-sectional dependence. This is done by generating the errors as
$$
\varepsilon_{it} = 0.5 \, \varepsilon_{(i-1)t} + \eta_{it} 
$$ 
for $\eta_{it} \sim N(0,1)$ independently in each wave of the panel, and the initial condition drawn from the implied steady-state distribution. Under this type of dependence  the bias again remains important as $\nicefrac{\N}{\T}\rightarrow c \in (0,+\infty)$.

\subsection{Baseline}
\begin{figure}[htbp] \centering 
\includegraphics[height=.97\textheight]{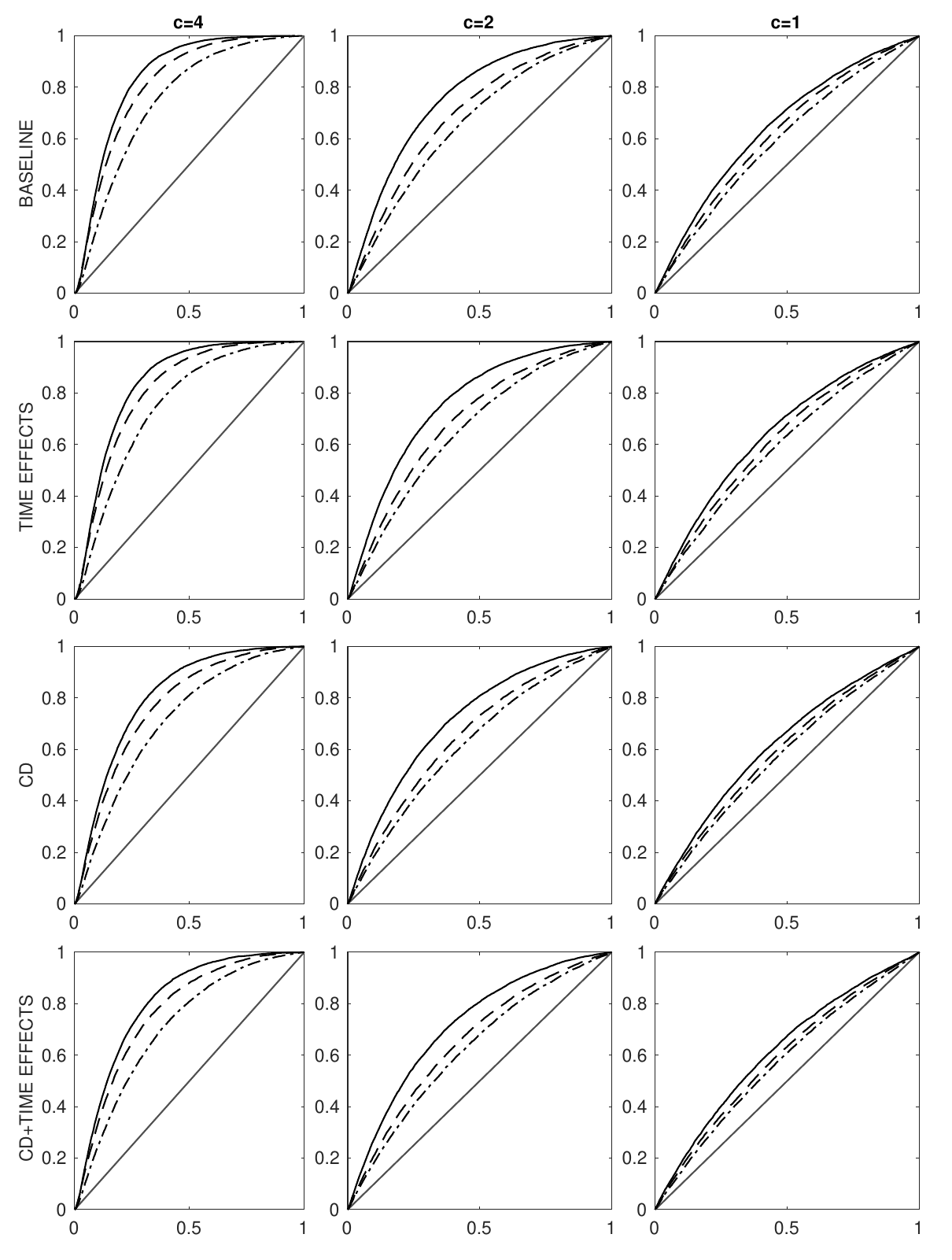} \caption{Empirical distribution of bootstrap $p$-values.} 
\label{fig:fig1} 
\end{figure}

We first illustrate Theorem \ref{thm:thm2} by plotting the empirical distribution function of the $p$-value
$$
\mathbb{P}^*(\hat{\beta}^* - \hat{\beta} \leq \hat{\beta} - {\beta} )
$$
for different ranges of $(\N,\T)$ and choices of $q$. The theorem implies that the $p$-value should be asymptotically uniform as $\nicefrac{\N}{\T}\rightarrow c \in [0,+\infty)$. The top three panels in Figure \ref{fig:fig1} plot the distribution for different values of $(\N,\T)$ so that either $c = 4$ (left plot), $c = 2$ (middle plot), or finally $c = 1$ (right plot). Each plot has $\N\in \lbrace 100, 200, 400\rbrace$. Hence, in the left-most plot we have $(\N,\T)=(100,25)$ (solid line), $(\N,\T)=(200,50)$ (dashed line), and $(\N,\T)=(400,100)$ (dashed-dotted line); in the middle plot we have $(\N,\T)=(100,50)$ (solid line), $(\N,\T)=(200,100)$ (dashed line), and $(\N,\T)=(400,200)$ (dashed-dotted line); and in the right-most plot we have $(\N,\T)=(100,100)$ (solid line), $(\N,\T)=(200,200)$ (dashed line), and $(\N,\T)=(400,400)$ (dashed-dotted line). Each plot includes the $45^\circ$ line as (asymptotic) reference. The block length was set as $q\in \lbrace 3,4,6,8,10 \rbrace$ for $\T\in \lbrace 25, 50, 100, 200, 400 \rbrace$, respectively, which is  in accordance with our theory. All the results were obtained over 10,000 Monte Carlo replications; the bootstrap uses 1,999 replications for each $q$.
As predicted, in each plot, as $\N,\T\rightarrow \infty$, the empirical distribution approaches the uniform distribution.

\begin{table}[p]
\caption{Simulation results for the baseline design}  \label{table:table1}
\centering 
\resizebox{.85\textwidth}{!}{%
\begin{tabular}{llrrrrrrrrrrrr}
\hline\hline
 & & WG & \multicolumn{5}{c}{PLUG-IN CORRECTION ($h$)} & SPJ & \multicolumn{5}{c}{BOOTSTRAP ($q$)} \\
\cline{3-3} \cline{4-8} \cline{9-9} \cline{10-14}
 &  & & $1$ & $3$ & $5$ & $7$ & $9$ & & $2$ & $4$ & $6$ & $8$ & $10$ \\
\hline
\multicolumn{14}{c}{ $(\N,\T) = (100, 25)$} \\
\hline
BIAS & & $-.0627$ & $-.0319$ & $-.0164$ & $-.0214$ & $-.0315$ & $-.0427$ &
$.0040$ & $-.0458$ & $-.0282$ & $-.0212$ & $-.0197$ & $-.0152$ \\

STD & & $.0185$ & $.0180$ & $.0188$ & $.0195$ & $.0197$ & $.0195$ &
$.0221$ & $.0185$ & $.0193$ & $.0202$ & $.0210$ & $.0219$ \\

SE ($h$) & $0$ & $.0180$ & $.0180$ & $.0180$ & $.0180$ & $.0180$ & $.0180$ &
$.0181$ & $.0180$ & $.0180$ & $.0180$ & $.0180$ & $.0181$ \\

& $2$ & $.0175$ & $.0172$ & $.0171$ & $.0172$ & $.0172$ & $.0173$ &
$.0171$ & $.0173$ & $.0172$ & $.0172$ & $.0172$ & $.0171$ \\

& $5$ & $.0172$ & $.0168$ & $.0167$ & $.0168$ & $.0168$ & $.0169$ &
$.0167$ & $.0170$ & $.0168$ & $.0168$ & $.0168$ & $.0168$ \\

SIZE ($h$)  & $0$ & $.9357$ & $.4214$ & $.1513$ & $.2373$ & $.4206$ & $.6481$ &
$.1144$ & $.7219$ & $.3487$ & $.2413$ & $.2285$ & $.1814$ \\

& $2$ & $.9413$ & $.4530$ & $.1788$ & $.2639$ & $.4519$ & $.6768$ &
$.1361$ & $.7445$ & $.3818$ & $.2649$ & $.2548$ & $.2089$ \\

& $5$ & $.9458$ & $.4705$ & $.1895$ & $.2762$ & $.4679$ & $.6895$ &
$.1460$ & $.7579$ & $.3973$ & $.2776$ & $.2682$ & $.2193$ \\

LENGTH ($h$) & $0$ & $.0705$ & $.0706$ & $.0707$ & $.0707$ & $.0706$ & $.0705$ &
$.0710$ & $.0705$ & $.0706$ & $.0707$ & $.0707$ & $.0708$ \\

& $2$ & $.0685$ & $.0675$ & $.0672$ & $.0673$ & $.0675$ & $.0678$ &
$.0669$ & $.0679$ & $.0674$ & $.0673$ & $.0673$ & $.0672$ \\

& $5$ & $.0673$ & $.0660$ & $.0657$ & $.0658$ & $.0660$ & $.0664$ &
$.0655$ & $.0665$ & $.0659$ & $.0658$ & $.0658$ & $.0657$ \\

SIZE (\%)& & \multicolumn{1}{c}{---}  & \multicolumn{1}{c}{---}  & \multicolumn{1}{c}{---}  & \multicolumn{1}{c}{---}  & \multicolumn{1}{c}{---}  & \multicolumn{1}{c}{---}  & \multicolumn{1}{c}{---}  & $.3638$ & $.0080$ & $.0011$ & $.0027$ & $.0083$ \\

LENGTH (\%) & & \multicolumn{1}{c}{---}  & \multicolumn{1}{c}{---}  & \multicolumn{1}{c}{---}  & \multicolumn{1}{c}{---}  & \multicolumn{1}{c}{---}  & \multicolumn{1}{c}{---}  & \multicolumn{1}{c}{---}  & $.0899$ & $.1204$ & $.1428$ & $.1348$ & $.1220$ \\

\hline
\multicolumn{14}{c}{$(\N,\T) = (100, 50)$} \\
\hline
BIAS & & $-.0309$ & $-.0156$ & $-.0059$ & $-.0055$ & $-.0074$ & $-.0099$ &
$.0011$ & $-.0227$ & $-.0142$ & $-.0105$ & $-.0088$ & $-.0077$ \\

STD & & $.0125$ & $.0123$ & $.0126$ & $.0128$ & $.0129$ & $.0129$ &
$.0136$ & $.0125$ & $.0127$ & $.0130$ & $.0133$ & $.0135$ \\

SE ($h$) & $0$ & $.0125$ & $.0125$ & $.0125$ & $.0125$ & $.0125$ & $.0125$ &
$.0125$ & $.0125$ & $.0125$ & $.0125$ & $.0125$ & $.0125$ \\

& $2$ & $.0123$ & $.0122$ & $.0121$ & $.0121$ & $.0122$ & $.0122$ &
$.0121$ & $.0123$ & $.0122$ & $.0122$ & $.0122$ & $.0122$ \\

& $5$ & $.0122$ & $.0120$ & $.0120$ & $.0119$ & $.0120$ & $.0120$ &
$.0119$ & $.0121$ & $.0120$ & $.0120$ & $.0120$ & $.0120$ \\

SIZE ($h$) & $0$ & $.6995$ & $.2307$ & $.0765$ & $.0781$ & $.0952$ & $.1292$ &
$.0759$ & $.4441$ & $.2060$ & $.1391$ & $.1215$ & $.1142$ \\

& $2$ & $.7080$ & $.2460$ & $.0854$ & $.0868$ & $.1059$ & $.1390$ &
$.0842$ & $.4557$ & $.2185$ & $.1521$ & $.1314$ & $.1242$ \\

& $5$ & $.7154$ & $.2559$ & $.0916$ & $.0918$ & $.1098$ & $.1450$ &
$.0909$ & $.4657$ & $.2258$ & $.1577$ & $.1377$ & $.1298$ \\

LENGTH ($h$) & $0$ & $.0489$ & $.0490$ & $.0490$ & $.0490$ & $.0490$ & $.0490$ &
$.0490$ & $.0489$ & $.0490$ & $.0490$ & $.0490$ & $.0490$ \\

& $2$ & $.0483$ & $.0479$ & $.0476$ & $.0476$ & $.0476$ & $.0477$ &
$.0475$ & $.0481$ & $.0478$ & $.0477$ & $.0477$ & $.0477$ \\

& $5$ & $.0478$ & $.0472$ & $.0469$ & $.0468$ & $.0469$ & $.0470$ &
$.0467$ & $.0474$ & $.0471$ & $.0470$ & $.0470$ & $.0469$ \\

SIZE (\%) & & \multicolumn{1}{c}{---}  & \multicolumn{1}{c}{---}  & \multicolumn{1}{c}{---}  & \multicolumn{1}{c}{---}  & \multicolumn{1}{c}{---}  & \multicolumn{1}{c}{---}  & \multicolumn{1}{c}{---}  & $.3480$ & $.0753$ & $.0228$ & $.0140$ & $.0129$ \\

LENGTH (\%)& & \multicolumn{1}{c}{---}  & \multicolumn{1}{c}{---}  & \multicolumn{1}{c}{---}  & \multicolumn{1}{c}{---}  & \multicolumn{1}{c}{---}  & \multicolumn{1}{c}{---}  & \multicolumn{1}{c}{---}  & $.0519$ & $.0576$ & $.0633$ & $.0677$ & $.0719$ \\

\hline
\multicolumn{14}{c}{$(\N,\T) = (100, 100)$} \\
\hline
BIAS & & $-.0153$ & $-.0077$ & $-.0025$ & $-.0016$ & $-.0019$ & $-.0024$ &
$.0002$ & $-.0114$ & $-.0071$ & $-.0051$ & $-.0041$ & $-.0035$ \\

STD & & $.0089$ & $.0088$ & $.0089$ & $.0089$ & $.0090$ & $.0090$ &
$.0093$ & $.0089$ & $.0089$ & $.0090$ & $.0091$ & $.0092$ \\

SE ($h$) & $0$ & $.0087$ & $.0087$ & $.0087$ & $.0087$ & $.0087$ & $.0087$ &
$.0087$ & $.0087$ & $.0087$ & $.0087$ & $.0087$ & $.0087$ \\

& $2$ & $.0087$ & $.0086$ & $.0086$ & $.0086$ & $.0086$ & $.0086$ &
$.0086$ & $.0087$ & $.0086$ & $.0086$ & $.0086$ & $.0086$ \\

& $5$ & $.0086$ & $.0086$ & $.0085$ & $.0085$ & $.0085$ & $.0085$ &
$.0085$ & $.0086$ & $.0086$ & $.0086$ & $.0086$ & $.0085$ \\

SIZE ($h$) & $0$ & $.4195$ & $.1405$ & $.0609$ & $.0602$ & $.0624$ & $.0653$ &
$.0658$ & $.2537$ & $.1287$ & $.0938$ & $.0829$ & $.0807$ \\

& $2$ & $.4260$ & $.1461$ & $.0649$ & $.0640$ & $.0664$ & $.0689$ &
$.0703$ & $.2594$ & $.1330$ & $.0979$ & $.0888$ & $.0848$ \\

& $5$ & $.4305$ & $.1496$ & $.0678$ & $.0672$ & $.0695$ & $.0714$ &
$.0725$ & $.2630$ & $.1372$ & $.1005$ & $.0918$ & $.0878$ \\

LENGTH ($h$) & $0$ & $.0343$ & $.0343$ & $.0343$ & $.0343$ & $.0343$ & $.0343$ &
$.0343$ & $.0343$ & $.0343$ & $.0343$ & $.0343$ & $.0343$ \\

& $2$ & $.0341$ & $.0339$ & $.0338$ & $.0338$ & $.0338$ & $.0338$ &
$.0337$ & $.0340$ & $.0339$ & $.0338$ & $.0338$ & $.0338$ \\

& $5$ & $.0339$ & $.0336$ & $.0335$ & $.0335$ & $.0335$ & $.0335$ &
$.0334$ & $.0337$ & $.0336$ & $.0336$ & $.0335$ & $.0335$ \\

SIZE (\%)  & & \multicolumn{1}{c}{---}  & \multicolumn{1}{c}{---}  &  \multicolumn{1}{c}{---} &  \multicolumn{1}{c}{---} &  \multicolumn{1}{c}{---} &  \multicolumn{1}{c}{---} & \multicolumn{1}{c}{---}  & $.2408$ & $.0979$ & $.0604$ & $.0501$ & $.0441$ \\

LENGTH (\%) & &  \multicolumn{1}{c}{---}   &  \multicolumn{1}{c}{---}  & \multicolumn{1}{c}{---}  & \multicolumn{1}{c}{---}  & \multicolumn{1}{c}{---}  & \multicolumn{1}{c}{---}  &  \multicolumn{1}{c}{---}  & $.0346$ & $.0355$ & $.0364$ & $.0373$ & $.0382$ \\

\hline\hline
\end{tabular}
}
\end{table}

Next, we turn to point estimation and inference. For brevity we focus on $\N=100$ and $\T\in \lbrace 25,50,100 \rbrace$. Table \ref{table:table1} contains the bias and standard deviation of the within-group (WG) estimator, the bias-corrected estimator of \cite{HahnKuersteiner2011} for different bandwidth choices $h\in \lbrace  1,3,5,7,9\rbrace$, for the split-panel jackknife (SPJ) estimator of \cite{DhaeneJochmans2015}, and for the bootstrap bias-corrected estimator as described above for $q\in \lbrace  2,4,6,8,10\rbrace$. All corrections reduce the bias in the within-group estimator. The ability of the analytical approach to correct for bias is quite sensitive to the bandwidth choice. The remaining bias in the bootstrap correction also varies with the block length, but somewhat less so. The jackknife is most successful at bias reduction but  also tends to be the most variable. 

The table also gives standard errors for each of these procedures using a HAC estimator of the asymptotic variance based on Bartlett weights and different bandwidths $h\in \lbrace 0,2,5\rbrace$, as well as the size of a two-sided $t$-test (at nominal level of $5\%$) and the average length of the implied $95\%$ confidence interval.
The actual variability of the various estimators is increasingly underestimated as the bandwidth in the HAC estimator grows. Consequently, the rejection frequency goes up whereas the average length of the confidence interval goes down with the bandwidth. Empirical size improves drastically as a consequence of bias correction, although the rejection frequency remains somewhat high in the shortest panels. 

Finally, the bottom two rows of each panel in the table also give the size and average confidence-interval length as obtained through the reverse-percentile bootstrap. Whereas inference based on the limit distribution tended to yield over-rejection under the null, the reverse-percentile bootstrap is conservative for larger block lengths, especially, again, in the shortest panels.



\begin{table}[p]
\caption{Simulation results for the estimator with time effects} \label{table:table2}
\centering 
\resizebox{.85\textwidth}{!}{%
\begin{tabular}{llrrrrrrrrrrrr}
\hline\hline
 & & WG & \multicolumn{5}{c}{PLUG-IN CORRECTION ($h$)} & SPJ & \multicolumn{5}{c}{BOOTSTRAP ($q$)} \\
\cline{3-3} \cline{4-8} \cline{9-9} \cline{10-14}
 &  & & $1$ & $3$ & $5$ & $7$ & $9$ & & $2$ & $4$ & $6$ & $8$ & $10$ \\
\hline

\multicolumn{14}{c}{$(\N,\T) = (100,25)$} \\ \hline
BIAS & & $-.0627$ & $-.0319$ & $-.0164$ & $-.0214$ & $-.0315$ & $-.0427$ & $.0040$ & $-.0458$ & $-.0281$ & $-.0212$ & $-.0197$ & $-.0152$ \\
STD & & $.0186$ & $.0181$ & $.0189$ & $.0196$ & $.0198$ & $.0197$ & $.0222$ & $.0186$ & $.0194$ & $.0203$ & $.0211$ & $.0221$ \\
SE ($h$)& 0 & $.0180$ & $.0180$ & $.0180$ & $.0180$ & $.0180$ & $.0180$ & $.0181$ & $.0180$ & $.0180$ & $.0180$ & $.0180$ & $.0181$ \\
 & 2 & $.0175$ & $.0172$ & $.0171$ & $.0172$ & $.0172$ & $.0173$ & $.0171$ & $.0173$ & $.0172$ & $.0172$ & $.0172$ & $.0171$ \\
& 5 & $.0172$ & $.0168$ & $.0167$ & $.0168$ & $.0168$ & $.0169$ & $.0167$ & $.0170$ & $.0168$ & $.0168$ & $.0168$ & $.0168$ \\
SIZE ($h$)  & 0 & $.9350$ & $.4225$ & $.1543$ & $.2372$ & $.4204$ & $.6486$ & $.1152$ & $.7187$ & $.3474$ & $.2420$ & $.2288$ & $.1849$ \\
& 2 & $.9392$ & $.4528$ & $.1802$ & $.2671$ & $.4521$ & $.6743$ & $.1376$ & $.7442$ & $.3810$ & $.2698$ & $.2580$ & $.2096$ \\
& 5 & $.9436$ & $.4707$ & $.1901$ & $.2788$ & $.4648$ & $.6885$ & $.1470$ & $.7568$ & $.3977$ & $.2811$ & $.2708$ & $.2216$ \\
LENGTH ($h$) & 0 & $.0705$ & $.0706$ & $.0707$ & $.0707$ & $.0706$ & $.0705$ & $.0710$ & $.0705$ & $.0706$ & $.0707$ & $.0707$ & $.0708$ \\
& 2 & $.0685$ & $.0675$ & $.0672$ & $.0673$ & $.0675$ & $.0678$ & $.0669$ & $.0679$ & $.0674$ & $.0673$ & $.0673$ & $.0672$ \\
 & 5 & $.0673$ & $.0660$ & $.0657$ & $.0658$ & $.0660$ & $.0664$ & $.0655$ & $.0665$ & $.0659$ & $.0658$ & $.0658$ & $.0657$ \\
SIZE (\%)  & & \multicolumn{1}{c}{---}  & \multicolumn{1}{c}{---}  &  \multicolumn{1}{c}{---} &  \multicolumn{1}{c}{---} &  \multicolumn{1}{c}{---} &  \multicolumn{1}{c}{---} & \multicolumn{1}{c}{---}  & $.3629$ & $.0086$ & $.0013$ & $.0027$ & $.0084$  \\
LENGTH (\%)  & & \multicolumn{1}{c}{---}  & \multicolumn{1}{c}{---}  &  \multicolumn{1}{c}{---} &  \multicolumn{1}{c}{---} &  \multicolumn{1}{c}{---} &  \multicolumn{1}{c}{---} & \multicolumn{1}{c}{---}  & $.0902$ & $.1207$ & $.1430$ & $.1349$ & $.1221$  \\
\hline
\multicolumn{14}{c}{$(\N,\T) = (100,50)$} \\ \hline
BIAS & & $-.0309$ & $-.0156$ & $-.0059$ & $-.0055$ & $-.0074$ & $-.0098$ & $.0011$ & $-.0227$ & $-.0142$ & $-.0105$ & $-.0088$ & $-.0077$ \\
STD & & $.0126$ & $.0124$ & $.0126$ & $.0128$ & $.0129$ & $.0130$ & $.0137$ & $.0126$ & $.0128$ & $.0130$ & $.0133$ & $.0136$ \\
SE ($h$) & 0 & $.0125$ & $.0125$ & $.0125$ & $.0125$ & $.0125$ & $.0125$ & $.0125$ & $.0125$ & $.0125$ & $.0125$ & $.0125$ & $.0125$ \\
 & 2 & $.0123$ & $.0122$ & $.0121$ & $.0121$ & $.0122$ & $.0122$ & $.0121$ & $.0123$ & $.0122$ & $.0122$ & $.0122$ & $.0122$ \\
 & 5 & $.0122$ & $.0120$ & $.0120$ & $.0119$ & $.0120$ & $.0120$ & $.0119$ & $.0121$ & $.0120$ & $.0120$ & $.0120$ & $.0120$ \\
SIZE ($h$) & 0 & $.6984$ & $.2311$ & $.0784$ & $.0790$ & $.0970$ & $.1287$ & $.0771$ & $.4448$ & $.2077$ & $.1392$ & $.1223$ & $.1155$ \\
 & 2 & $.7079$ & $.2469$ & $.0872$ & $.0880$ & $.1073$ & $.1388$ & $.0870$ & $.4590$ & $.2190$ & $.1514$ & $.1318$ & $.1250$ \\
& 5 & $.7155$ & $.2555$ & $.0921$ & $.0934$ & $.1119$ & $.1447$ & $.0922$ & $.4680$ & $.2277$ & $.1575$ & $.1382$ & $.1307$ \\
LENGTH ($h$) & 0 & $.0489$ & $.0490$ & $.0490$ & $.0490$ & $.0490$ & $.0490$ & $.0490$ & $.0489$ & $.0490$ & $.0490$ & $.0490$ & $.0490$ \\
& 2 & $.0483$ & $.0479$ & $.0476$ & $.0476$ & $.0476$ & $.0477$ & $.0475$ & $.0481$ & $.0478$ & $.0477$ & $.0477$ & $.0477$ \\
& 5 & $.0478$ & $.0472$ & $.0469$ & $.0468$ & $.0469$ & $.0470$ & $.0467$ & $.0474$ & $.0471$ & $.0470$ & $.0470$ & $.0469$ \\
SIZE (\%)  & & \multicolumn{1}{c}{---}  & \multicolumn{1}{c}{---}  &  \multicolumn{1}{c}{---} &  \multicolumn{1}{c}{---} &  \multicolumn{1}{c}{---} &  \multicolumn{1}{c}{---} & \multicolumn{1}{c}{---}  & $.3468$ & $.0759$ & $.0241$ & $.0139$ & $.0128$  \\
LENGTH (\%)  & & \multicolumn{1}{c}{---}  & \multicolumn{1}{c}{---}  &  \multicolumn{1}{c}{---} &  \multicolumn{1}{c}{---} &  \multicolumn{1}{c}{---} &  \multicolumn{1}{c}{---} & \multicolumn{1}{c}{---} & $.0521$ & $.0578$ & $.0635$ & $.0679$ & $.0721$  \\
\hline
\multicolumn{14}{c}{$(\N,\T) = (100,100)$} \\ \hline
BIAS & & $-.0153$ & $-.0077$ & $-.0025$ & $-.0016$ & $-.0019$ & $-.0024$ & $.0002$ & $-.0114$ & $-.0071$ & $-.0051$ & $-.0041$ & $-.0035$ \\
STD & & $.0089$ & $.0088$ & $.0089$ & $.0090$ & $.0090$ & $.0090$ & $.0093$ & $.0089$ & $.0090$ & $.0091$ & $.0091$ & $.0092$ \\
SE ($h$) & 0 & $.0087$ & $.0087$ & $.0087$ & $.0087$ & $.0087$ & $.0087$ & $.0087$ & $.0087$ & $.0087$ & $.0087$ & $.0087$ & $.0087$ \\
 & 2 & $.0087$ & $.0086$ & $.0086$ & $.0086$ & $.0086$ & $.0086$ & $.0086$ & $.0087$ & $.0086$ & $.0086$ & $.0086$ & $.0086$ \\
 & 5 & $.0086$ & $.0086$ & $.0085$ & $.0085$ & $.0085$ & $.0085$ & $.0085$ & $.0086$ & $.0086$ & $.0086$ & $.0086$ & $.0085$ \\
SIZE ($h$) & 0 & $.4198$ & $.1415$ & $.0616$ & $.0591$ & $.0616$ & $.0662$ & $.0666$ & $.2560$ & $.1303$ & $.0941$ & $.0846$ & $.0808$ \\
 & 2 & $.4241$ & $.1464$ & $.0666$ & $.0627$ & $.0662$ & $.0706$ & $.0693$ & $.2612$ & $.1351$ & $.0989$ & $.0889$ & $.0857$ \\
 & 5 & $.4284$ & $.1508$ & $.0695$ & $.0668$ & $.0689$ & $.0736$ & $.0711$ & $.2667$ & $.1384$ & $.1021$ & $.0914$ & $.0879$ \\
LENGTH ($h$) & 0 & $.0343$ & $.0343$ & $.0343$ & $.0343$ & $.0343$ & $.0343$ & $.0343$ & $.0343$ & $.0343$ & $.0343$ & $.0343$ & $.0343$ \\
& 2 & $.0341$ & $.0339$ & $.0338$ & $.0338$ & $.0338$ & $.0338$ & $.0337$ & $.0340$ & $.0339$ & $.0338$ & $.0338$ & $.0338$ \\
& 5 & $.0339$ & $.0336$ & $.0335$ & $.0335$ & $.0335$ & $.0335$ & $.0334$ & $.0337$ & $.0336$ & $.0336$ & $.0335$ & $.0335$ \\
SIZE (\%)  & & \multicolumn{1}{c}{---}  & \multicolumn{1}{c}{---}  &  \multicolumn{1}{c}{---} &  \multicolumn{1}{c}{---} &  \multicolumn{1}{c}{---} &  \multicolumn{1}{c}{---} & \multicolumn{1}{c}{---}  & $.2365$ & $.0983$ & $.0619$ & $.0504$ & $.0443$  \\
LENGTH (\%)  & & \multicolumn{1}{c}{---}  & \multicolumn{1}{c}{---}  &  \multicolumn{1}{c}{---} &  \multicolumn{1}{c}{---} &  \multicolumn{1}{c}{---} &  \multicolumn{1}{c}{---} & \multicolumn{1}{c}{---} & $.0348$ & $.0356$ & $.0366$ & $.0374$ & $.0383$  \\
\hline\hline
\end{tabular}%
}
\end{table}

\subsection{Extensions}
The second row of plots in Figure \ref{fig:fig1} contains the empirical distributions of the $p$-values when time effects are estimated along with unit-specific effects. Table \ref{table:table2} provides the corresponding results on estimation and inference and mimics Table \ref{table:table1} in layout. We clarify that only the least-squares estimator changes relative to the baseline case. None of the bias corrections were implemented differently, and neither was the reverse-percentile method. The results are essentially the same as in the baseline case, thus demonstrating that the bootstrap (and, indeed, the other approaches considered) is robust to the inclusion of time effects.


\begin{table}[h!]
\caption{Simulation results with cross-sectional dependence (CD)}  \label{table:table3}
\centering
\resizebox{.70\textwidth}{!}{%
\begin{tabular}{lcccccccccc}
\hline\hline
	& \multicolumn{10}{c}{BOOTSTRAP ($q$)}
	 \\
	 \hline
	& 2 & 4 & 6 & 8 & 10 & 2 & 4 & 6 & 8 & 10 \\
	\hline
 & \multicolumn{5}{c}{CD} & \multicolumn{5}{c}{CD + TIME EFFECTS} \\
 \hline
	  \multicolumn{11}{c}{$(\N,\T) = (100, 25)$} \\
	\hline
SIZE (\%)
	& $.2863$ & $.0240$ & $.0072$ & $.0117$ & $.0289$
	& $.2859$ & $.0230$ & $.0072$ & $.0134$ & $.0286$ \\
LENGTH (\%)
	& $.1058$ & $.1318$ & $.1510$ & $.1439$ & $.1309$
	& $.1061$ & $.1320$ & $.1511$ & $.1441$ & $.1311$ \\
	\hline
	  \multicolumn{11}{c}{$(\N,\T) = (100, 50)$} \\
	\hline
SIZE (\%)
	& $.2476$ & $.0743$ & $.0371$ & $.0293$ & $.0304$
	& $.2471$ & $.0750$ & $.0387$ & $.0309$ & $.0295$ \\
LENGTH (\%)
	& $.0650$ & $.0691$ & $.0734$ & $.0768$ & $.0800$
	& $.0651$ & $.0692$ & $.0735$ & $.0769$ & $.0801$ \\
	\hline
	  \multicolumn{11}{c}{$(\N,\T) = (100, 100)$} \\
	\hline
SIZE (\%)
	& $.1644$ & $.0868$ & $.0687$ & $.0632$ & $.0623$
	& $.1671$ & $.0861$ & $.0682$ & $.0632$ & $.0627$ \\
LENGTH (\%)
	& $.0442$ & $.0447$ & $.0452$ & $.0456$ & $.0461$
	& $.0443$ & $.0448$ & $.0453$ & $.0457$ & $.0462$ \\
	\hline\hline
\end{tabular}%
}
\end{table}

Our remaining simulation results pertain to data with cross-sectional dependence. The third row of Figure \ref{fig:fig1} plots the $p$-value distributions based on the within-group estimator in the presence of cross-sectional dependence. The fourth row does the same when the estimator used accounts for time effects. 
Table \ref{table:table3} provides the corresponding inference results for the reverse-percentile bootstrap. The HAC variance estimator is no longer valid here (as it ignores the presence of cross-sectional dependence) and so results based on it are not included here. In either case, both the figure and the table illustrate the robustness of the bootstrap procedure.

Whereas, in our design, the presence of both cross-sectional dependence and time effects causes no complications, we remark that this need not always be the case. In general, the regressors and the errors may each be correlated across
units, and the regressors of one unit may be correlated with the errors of another unit in diﬀerent time periods. If, however, the regressors of one unit are correlated with the errors of other units in the same time period, the consequences depend on the particular structure of the dependence, with it being possible that the bootstrap fails to replicate all relevant asymptotic bias. We leave this for future work.




\section{Concluding remarks}
The within-group estimator of linear models computed from $\N\times \T$ panel data suffers from asymptotic bias under asymptotics where $\nicefrac{\N}{\T} \rightarrow c \in (0,+\infty)$ when regressors are not strictly exogenous. We have shown that, in spite of the presence of this bias, the moving block bootstrap permits valid inference based on the within-group estimator. This finding presents an encouraging addition to the literature where, until now, it was not known whether models with unspecified feedback were amenable to bootstrapping in the presence of incidental-parameter bias. 
The approach should extend naturally to more general nonlinear models.

We have no theoretical argument to prefer the bootstrap over the various bias-correction approaches, in the
same way that those different approaches themselves cannot be ranked under our asymptotic
framework. A practical advantage of the bootstrap, though, is that it makes inference straightforward, and that 
it does not require explicit estimators for the bias and variance.

The bootstrap requires a tuning parameter in the form of the block length. Analytical and jackknife approaches, in turn, require choosing bandwidths and kernel functions. We have gauged the sensitivity of the various methods to such choices in the simulations. However, little is known about how to choose such parameters in an optimal manner in the panel data case, due to the presence of bias. This presents an interesting avenue for further research.



\appendix
\renewcommand{\theequation}{A.\arabic{equation}} 
\setcounter{equation}{0}
\renewcommand{\thelemma}{A.\arabic{lemma}} 
\setcounter{lemma}{0}
\section*{Appendix} 
For notational simplicity, and without loss of generality, we take $x_{it}$ to be univariate throughout the appendix. 

\subsection*{Proof of Theorem \ref{thm:thm1}}
From the definition of $\hat{\beta}$, 
$$
\sqrt{\N\T} (\hat{\beta}-\beta)
=
\hat{\varSigma}_{\N,\T}^{-1} 
\left(
\frac{1}{\sqrt{\N\T}} \sum_{i=1}^{\N} \sum_{t=1}^{\T} (x_{it} - \bar{x}_i) \, \varepsilon_{it}
\right)
,
$$
where, recall,  
$
\hat{\varSigma}_{\N,\T}
= 
\nicefrac{1}{\N\T}
\sum_{i=1}^{\N} \sum_{t=1}^{\T} (x_{it} - \bar{x}_i )^2. 
$

We first show that $\lvert \hat{\varSigma}_{\N,\T} - {\varSigma}_{\N,\T} \rvert  = o_P(1)$. By using the triangle inequality we can write
\begin{equation} \label{eq:jacobian1}
\lvert \hat{\varSigma}_{\N,\T} - {\varSigma}_{\N,\T} \rvert  
\leq
\lvert \hat{\varSigma}_{\N,\T} - \check{\varSigma}_{\N,\T} \rvert  
+
\lvert \check{\varSigma}_{\N,\T} - {\varSigma}_{\N,\T} \rvert ,
\end{equation}
for 
$$
\check{\varSigma}_{\N,\T}
\coloneqq
\frac{1}{\N\T}
\sum_{i=1}^{\N} \sum_{t=1}^{\T} (x_{it} - \mathbb{E}({x}_i) )^{2} 
=
\frac{1}{\N\T}
\sum_{i=1}^{\N} \sum_{t=1}^{\T} z_{it}^{2}. 
$$
We proceed by showing that each of the terms on the right-hand side converges to zero in probability. 

For the first term in \eqref{eq:jacobian1}, on working out the square and re-arranging we observe that
$$
\hat{\varSigma}_{\N,\T} - \check{\varSigma}_{\N,\T}
=
- \frac{1}{\N} \sum_{i=1}^{\N} \bar{z}_i^2, 
$$
with $\bar{z}_i \coloneqq \nicefrac{1}{\T} \sum_{t=1}^{\T} z_{it}$. 
Therefore, for any $\epsilon>0$,
$$
\mathbb{P}
(
\lvert 
\hat{\varSigma}_{\N,\T} - \check{\varSigma}_{\N,\T}
\rvert
> \epsilon
)
\leq
\frac{\mathbb{E}(\lvert 
\hat{\varSigma}_{\N,\T} - \check{\varSigma}_{\N,\T}
\rvert)}{\epsilon}
=
\frac{\nicefrac{1}{\N} \sum_{i=1}^{\N}
\mathbb{E}( \bar{z}_i ^2)}{\epsilon}
=
O(\T^{-1})
$$
by an application of Markov's inequality and the fact that
$
\sup_{i}\mathbb{E}( \bar{z}_i ^2) = O(\T^{-1}).
$
The latter claim follows from \citet[Corollary 3]{Hansen1991}.  By Assumption \ref{ass:ass1} the time-series processes $\lbrace z_{it} \rbrace$ are strong mixing with mixing coefficients $a_i$ for which $\sup_{i}a_i(h) \eqqcolon \bar{a}(h)$ satisfies the summability condition
$
 \sum_{h=1}^{+\infty} \bar{a}(h)^{\nicefrac{1}{2} - \nicefrac{1}{r}} < +\infty
$
for some $r>2$ for which $ \sup_{i}\mathbb{E}(\lvert z_{it }\rvert^r)$ is bounded. Therefore, by an application of \citeauthor{Hansen1991}'s (\citeyear{Hansen1991}) corollary, 
$$
\textstyle
\sup_{i}
\mathbb{E}\left(\left( \sum_{t=1}^{\T}z_{it} \right)^2\right)
\lesssim
\sum_{t=1}^{\T} 
\sup_{i}
\mathbb{E}(\lvert z_{it }\rvert^r)^{\nicefrac{2}{r}}
=
O(\T),
$$
where, here and later, we use $A\lesssim B$ to indicate that there exists a finite constant $C$ such that $A \leq C\, B$. From this it then follows that $
\sup_{i}\mathbb{E}( \bar{z}_i^2) = O(\T^{-1}),
$
as claimed. This handles the first term in \eqref{eq:jacobian1}.

For the second term in \eqref{eq:jacobian1}, we have
$$
\check{\varSigma}_{\N,\T}
-
{\varSigma}_{\N,\T}
=
\frac{1}{\N\T} \sum_{i=1}^{\N} \sum_{t=1}^{\T}
(
z_{it}^2
-
\mathbb{E}(z_{it}^2)
).
$$
Therefore, for any $\epsilon>0$,
$$
\mathbb{P}
\left(
\left\lvert
\check{\varSigma}_{\N,\T}
-
{\varSigma}_{\N,\T}
\right\rvert > \epsilon
\right)
\leq 
\frac{\nicefrac{1}{\N^2} \sum_{i=1}^{\N} \mathbb{E}(\left( \nicefrac{1}{\T} \sum_{t=1}^{\T} (z_{it}^2 - \mathbb{E}(z_{it}^2)) \right)^2)}{\epsilon^2} = O(\N^{-1} \T^{-1}),
$$
by an application of Chebychev's inequality and the fact that, by another application of \citet[Corollary 3]{Hansen1991},
$$
\sup_{i} 
\mathbb{E}
\left( 
\left(
\sum_{t=1}^{\T} (z_{it}^2 - \mathbb{E}(z_{it}^2)) \right)^2 \right) = O(\T)
$$
follows in the same way as before. 
With both right-hand side terms of \eqref{eq:jacobian1} $o_P(1)$ we obtain the desired result
 that $\lvert \hat{\varSigma}_{\N,\T} - \varSigma_{\N,\T}\rvert = o_P(1)$.

We next derive the limit distribution of the within-group least-squares normal equations,
$$
\frac{1}{\sqrt{\N\T}} \sum_{i=1}^{\N} \sum_{t=1}^{\T}
(x_{it} - \bar{x}_i) \, \varepsilon_{it} 
.
$$
Adding and subtracting $\mathbb{E}(x_{it}) (\varepsilon_{it} - \bar{\varepsilon}_i)$ to the summands and re-arranging allows us to write this as
\begin{equation} \label{eq:decomposition}
\frac{1}{\sqrt{\N\T}}
\sum_{i=1}^{\N} \sum_{t=1}^{\T}
z_{it} \varepsilon_{it} 
-
\left(\sqrt{\frac{\N}{\T}} \right)  \, \frac{1}{\N}
\sum_{i=1}^{\N} 
{\left(\frac{1}{\sqrt{\T}}\sum_{t=1}^{\T} z_{it} \right) \left( \frac{1}{\sqrt{\T}} \sum_{t=1}^{\T} \varepsilon_{it} \right)}
,
\end{equation}
and we analyse each of the terms in turn.

The first term in \eqref{eq:decomposition} is a scaled sample average of zero-mean random variables with variance covariance matrix $\varOmega_{\N,\T}$ that satisfies the conditions of \citet[Theorem 5.20]{White1984} (with the required order of mixing following from Assumption \ref{ass:ass1} together with Theorem 3.49 of \cite{White1984}). Therefore,  it converges in law to a normal random variable with mean zero and variance $\lim_{\N,\T\rightarrow\infty} \varOmega_{\N,\T}$.

The second term in \eqref{eq:decomposition}, on the other hand, will generate bias. To see this we introduce
$$
\chi_i \coloneqq
\left(\frac{1}{\sqrt{\T}}\sum_{t=1}^{\T} z_{it} \right) \left( \frac{1}{\sqrt{\T}} \sum_{t=1}^{\T} \varepsilon_{it} \right)
=
\frac{1}{\T} \sum_{t_1=1}^{\T} \sum_{t_2=1}^{\T} z_{it_1} \varepsilon_{it_2}.
$$
Recalling that $\mathbb{E}(z_{it} \varepsilon_{it}) = 0$, we have
$$
\mathbb{E}(\chi_i)
=
\sum_{\tau=1}^{\T-1} 
\left( \frac{\T-\tau}{\T} \right)
\left(
\mathbb{E}(z_{it} \varepsilon_{it+\tau})
+
\mathbb{E}(z_{it+\tau} \varepsilon_{it})
\right),
$$
and so
$
-
\nicefrac{1}{\N} \sum_{i=1}^{\N} \mathbb{E}(\chi_i ) =   b_{\N,\T}.
$
We now show that
\begin{equation} \label{eq:biasplim}
\frac{1}{\N} \sum_{i=1}^{\N} (\chi_i - \mathbb{E}(\chi_i)) = o_P(1).
\end{equation}
For any $\epsilon>0$, by Chebychev's inequality and independence of the data in the cross-section,
$$
\mathbb{P}\left(\left\lvert \frac{1}{\N} \sum_{i=1}^{\N} (\chi_i - \mathbb{E}(\chi_i)) \right\rvert > \epsilon \right)
\leq
\frac{1}{\N}
\frac{\nicefrac{1}{\N} \sum_{i=1}^{\N} \mathbb{E}((\chi_i - \mathbb{E}(\chi_i))^2)}{\epsilon^2}
\leq \frac{1}{\N} \frac{\sup_{i}\mathrm{var}(\chi_i) }{\epsilon^2}
,
$$
so that it suffices to show that $\sup_{i}\mathrm{var}(\chi_i) = O(1)$. Note that the variance of $\T\chi_i$ equals
\begin{equation} \label{eq:cs}
\mathbb{E}
\left(
\left( \sum_{t_1=1}^{\T} z_{it_1} \right)^2
\left( \sum_{t_2=1}^{\T} \varepsilon_{it_2} \right)^2
\right)
-
\mathbb{E}\left(
\sum_{t_1=1}^{\T}  \sum_{t_2=1}^{\T} z_{it_1} \varepsilon_{it_2}
\right)
\mathbb{E}\left(
\sum_{t_1=1}^{\T}  \sum_{t_2=1}^{\T} z_{it_1} \varepsilon_{it_2}
\right).
\end{equation}
For the first term in \eqref{eq:cs}, the Cauchy-Schwarz inequality yields
\begin{equation*}
\begin{split}
\mathbb{E}
\left(
\left( \sum_{t_1=1}^{\T} z_{it_1} \right)^2
\left( \sum_{t_2=1}^{\T} \varepsilon_{it_2} \right)^2
\right)
& \leq 
\sqrt{\mathbb{E}\left( \left( \sum_{t_1=1}^{\T} z_{it_1} \right)^4 \right)}
\sqrt{\mathbb{E}\left( \left( \sum_{t_2=1}^{\T} \varepsilon_{it_2} \right)^4 \right)},
\end{split}
\end{equation*}
while, by an application of \citet[Corollary 3]{Hansen1991}, 
\begin{equation*}
\begin{split}
\textstyle
\sup_{i}
\mathbb{E}\left(\left( \sum_{t=1}^{\T}z_{it} \right)^4\right)^{\nicefrac{1}{4}}
&
\lesssim
\sqrt{
\sum_{t=1}^{\T} 
\sup_{i}
\mathbb{E}(\lvert z_{it }\rvert^{2r})^{\nicefrac{1}{r}}}
=
O(\sqrt{\T}), 
\\
\textstyle
\sup_{i}
\mathbb{E}\left(\left( \sum_{t=1}^{\T} \varepsilon_{it} \right)^4\right)^{\nicefrac{1}{4}}
&
\lesssim
\sqrt{
\sum_{t=1}^{\T} 
\sup_{i}
\mathbb{E}(\lvert \varepsilon_{it }\rvert^{2r})^{\nicefrac{1}{r}}}
=
O(\sqrt{\T}), 
\end{split}
\end{equation*}
so that
$
\sup_{i}
\mathbb{E}
\left(
\left( \sum_{t_1=1}^{\T} z_{it_1} \right)^2
\left( \sum_{t_2=1}^{\T} \varepsilon_{it_2} \right)^2
\right)
=
O(\T^2).
$
For the second term in \eqref{eq:cs}, we have
$$
\sup_{i}
\mathbb{E}\left(
\left\lvert
\sum_{t_1=1}^{\T}  \sum_{t_2=1}^{\T} z_{it_1} \varepsilon_{it_2}
\right\rvert
\right)
\leq
\sqrt{\sup_{i} \mathbb{E}\left(
\left\lvert
\sum_{t=1}^{\T}   z_{it}
\right\rvert^2
\right)}
\,
\sqrt{ \sup_{i}
\mathbb{E}\left(
\left\lvert
\sum_{t=1}^{\T}   \varepsilon_{it}
\right\rvert^2
\right)
}
=
O(\T),
$$
with the order of magnitude already having been established above. It thus follows that the term in \eqref{eq:cs} is $O(\T^2)$ uniformly, and so that $\sup_{i} \mathrm{var}(\chi_i) = O(1)$. Therefore \eqref{eq:biasplim} holds.
Finally,  by Davydov's inequality (see, e.g., \citealt[Corollary A.2]{HallHeyde1980}), we have that
$$
\sum_{\tau=1}^{\infty} \lvert \mathbb{E}( z_{it} \varepsilon_{it+\tau} ) \rvert
\leq
8 \, \sup_{i,t} \mathbb{E}( \lvert z_{it} \rvert^r )^{\nicefrac{1}{r}} \, \sup_{i,t} \mathbb{E}( \lvert \varepsilon_{it} \rvert^r )^{\nicefrac{1}{r}} \sum_{\tau=1}^{\infty} \bar{a}( \tau )^{1-\nicefrac{2}{r}}
\leq
C, 
$$
where, here and later, $C$ indicates a finite constant whose precise value may be different on different occasions, and we have used the moment and mixing conditions in Assumption \ref{ass:ass1}. In the same way,  $\sum_{\tau=1}^{\infty} \lvert \mathbb{E}( z_{it+\tau} \varepsilon_{it} ) \rvert \leq C$, from which, as $0 \leq \nicefrac{( \T-\tau )}{\T} \leq 1$, it then follows that $b_{\N,\T} = O( 1 )$. 

Combining all results obtained reveals that
$$
\sqrt{\N\T} (\hat{\beta}-\beta)
=
\frac{1}{\sqrt{\N\T}} \sum_{i=1}^{\N} \sum_{t=1}^{\T}  {\varSigma}_{\N,\T}^{-1} z_{it}^{\vphantom{-1}} \varepsilon_{it}
+
\left(\sqrt{\frac{\N}{\T}} \right)  \, 
{\varSigma}_{\N,\T}^{-1} b_{\N,\T}^{\vphantom{-1}}
+
o_P(1),
$$
from which the theorem follows readily. \qed

\subsection*{Proof of Theorem \ref{thm:thm2}}

We begin by stating two identities concerning bootstrap expectations which we will make use of repeatedly. 

\begin{lemma} \label{lemma:Estar}
Let $\varpi_1,\ldots, \varpi_p$ be independent uniformly distributed on $\lbrace 0,1,\ldots,\T-\q \rbrace$ and let $\mathbb{E}^*$ denote expectations taken with respect to $\varpi_1,\ldots, \varpi_p$. For any function $f$ we then have
\begin{equation*} 
\mathbb{E}^*(f(\varpi_{p^\prime})) = \frac{1}{( \T-\q+1 )} \sum_{t=0}^{\T-\q} f( t ),
\qquad 1 \leq p^\prime \leq p.
\end{equation*}
Further, 
\begin{equation*} 
\mathbb{E}^*\left( \left\lvert \frac{1}{\p} \sum_{\p^\prime=1}^{\p} f( \varpi_{\p^\prime} ) \right\rvert^2 \right)
=
\frac{1}{\p} \, \mathbb{E}^*( \lvert f( \varpi_{p^\prime} ) \rvert^2 )
=
\frac{1}{p( \T-\q+1 )} \sum_{t=0}^{\T-\q} f( t )^2,
\end{equation*}
whenever $\mathbb{E}^*( f( \varpi_{p^\prime} ) ) = 0$.
\end{lemma}

\begin{proof}
For each $1\leq p^\prime \leq p$, the variable $\varpi_{p^\prime}$ takes on each value $t$ in $\lbrace 0,1,\ldots,\T-\q \rbrace$ with probability $\nicefrac{1}{( \T-\q+1 )}$. Hence, 
$$
\mathbb{E}^*( f( \varpi_{p^\prime} ) )
=
\sum_{t=0}^{\T-\q} f( t ) \, \mathbb{P}^*( \varpi_{p^\prime} = t )
=
\frac{1}{( \T-\q+1 )} \sum_{t=0}^{\T-\q} f( t ),
$$
which is the first identity in the lemma. Next, because the $\varpi_{p^\prime}$ are independent, it holds that
$$
\mathbb{E}^*\left( \left\lvert \frac{1}{\p} \sum_{\p^\prime=1}^{\p} f( \varpi_{\p^\prime} ) \right\rvert^2 \right)
=
\frac{1}{\p^2} \sum_{\p^\prime=1}^{\p} \mathbb{E}^*( f( \varpi_{\p^\prime} )^2 )
+
\frac{1}{\p^2} \sum_{\p^\prime=1}^{\p} \sum_{\p^{\prime\prime} \neq \p^\prime} \mathbb{E}^*( f( \varpi_{\p^\prime} ) ) \, \mathbb{E}^*( f( \varpi_{\p^{\prime\prime}} ) ) ,
$$
and the second identity follows from the fact that $\mathbb{E}^*( f( \varpi_{p^\prime} ) ) = 0$ and applying the first identity.
\end{proof}

We next state and prove two lemmas that will be used at various parts throughout the main proof.

\begin{lemma} \label{lemma:wsum}
Let $\lbrace w_{it} \rbrace_t$ be a mean zero strong-mixing sequence for each $i$ and assume that

\medskip\noindent
(i)  $\sup_i\sup_{t} \mathbb{E}( \lvert w_{it} \rvert^{ar} )$ is bounded for some $r>2$ and $a \geq 1$,

\medskip\noindent
(ii) the mixing coefficients are uniformly bounded by $\bar{a}( h ) = O( h^{-s} )$ for some $s  > \nicefrac{2ar}{(r-2)}$.

\medskip\noindent
Let $q \leq \nicefrac{m}{2}$ and write
$$
\bar{w}_i \coloneqq \frac{1}{\T} \sum_{t=1}^{\T} w_{it},
\qquad
\bar{w}_i^* \coloneqq \frac{1}{\T} \sum_{t=1}^{\T} w_{it}^*,
$$
with
$w_{i \, (\p^\prime-1)\q+\q^\prime}^* \coloneqq w_{i \varpi_{\p^\prime} + \q^\prime}$ for $1 \leq \p^\prime \leq \p$ and $1 \leq \q^\prime \leq \q$. Then, for all $0 \leq t \leq \T-\ell$ and each $1 \leq \ell \leq \T$, 

\medskip\noindent
(a) $\mathbb{E}( \lvert \sum_{l=1}^{\ell} w_{it+l} \rvert^{2a} ) \leq C \ell^{a}$ for $1 \leq \ell \leq \T$ and $0 \leq t \leq \T-\ell$,

\medskip\noindent
(b) $\mathbb{E}( \lvert \mathbb{E}^*( \bar{w}_i^* ) \rvert^{2a} ) \leq {C}{\T^{-a}}$, and

\medskip\noindent
(c) $\mathbb{E}( \lvert \bar{w}_i - \mathbb{E}^*( \bar{w}_i^* ) \rvert^{2} ) \leq C \, \q \T^{-2}$,

\medskip\noindent
where  $C$ is a finite constant that does not depend on $i$, $t$, $\ell$, $\q$, or $\T$.
\end{lemma}

\begin{proof}
Let $a_i( h )$ be the mixing coefficients of the sequence $\lbrace w_{it} \rbrace$. Then, by Condition (ii), 
$$
	\sup_{i} \sum_{h=1}^{\infty} a_i( h )^{\nicefrac{1}{2a}-\nicefrac{1}{ar}}
	\leq
	\sum_{h=1}^{\infty} \bar{a}( h )^{\nicefrac{(r-2)}{2ar}}
	\leq
	C \sum_{h=1}^{\infty} h^{-\nicefrac{s(r-2)}{2ar}}
	<
	\infty .
$$
An application of \citet[Corollary 3]{Hansen1991} gives 
\begin{equation*}
\begin{split}
\mathbb{E}\left( \left\lvert \sum_{l=1}^{\ell} w_{it+l} \right\rvert^{2a} \right)
& \leq
\mathbb{E}\left( \max_{1\leq k \leq \ell} \left\lvert \sum_{l=1}^k w_{it+l} \right\rvert^{2a} \right)
\\
& \leq
C 
\left(
\sum_{l=1}^\ell \mathbb{E}(\lvert w_{it+l}\rvert^{ar})^{\nicefrac{2}{ar}} 
\right)^a
\leq
C \, \ell^a \, \sup_{i} \sup_t \mathbb{E}(\lvert w_{it} \rvert^{ar})^{\nicefrac{2}{r}} ,
\end{split}
\end{equation*}
where the constant $C$ does not depend on $i$, $t$, or $\ell$. This proves Part (a) of the lemma.

Next, 
$$
\mathbb{E}^*( \bar{w}_i^* )
=
\frac{1}{\T} \sum_{\p^\prime=1}^{\p} \sum_{\q^\prime=1}^{\q} \mathbb{E}^*\big( w_{i\varpi_{\p^\prime}+\q^\prime} \big)
=
\frac{1}{\q} \sum_{\q^\prime=1}^{\q} \left( \frac{1}{( \T-\q+1 )} \sum_{t=0}^{\T-\q} w_{it+\q^\prime} \right),
$$
using Lemma \ref{lemma:Estar} and recalling that $\T = \p \q$. Then, using Jensen's inequality in a first step and Part (a) in the second step, 
$$
\mathbb{E}\left( \lvert \mathbb{E}^*( \bar{w}_i^* ) \rvert^{2a} \right)
\leq
\frac{1}{\q} \sum_{\q^\prime=1}^{\q} \frac{1}{( \T-\q+1 )^{2a}} \, \mathbb{E}\left( \left\lvert \sum_{t=0}^{\T-\q} w_{it+\q^\prime} \right\rvert^{2a} \right)
\leq
 \frac{C}{( \T-\q+1 )^{a}}
\leq 
C \, \left(\frac{2}{\T} \right)^a; 
$$
the last transition exploits $q \leq \nicefrac{\T}{2}$. This proves Part (b) of the lemma.

For the third part, finally, we begin by 
using that
$$
\mathbb{E}^*( \bar{w}_i^* )
=
\frac{1}{\q} \sum_{\q^\prime=1}^{\q} \frac{1}{( \T-\q+1 )} \sum_{t=0}^{\T-\q} w_{it+\q^\prime}
$$
by Lemma \ref{lemma:Estar}, to write
\begin{equation*}
\begin{split}
\bar{w}_i - \mathbb{E}^*( \bar{w}_i^* )
& =
\frac{1}{\q} \sum_{\q^\prime=1}^{\q} 
	\left(
	\frac{1}{\T} \sum^{\T}_{t=1} w_{it}
	-
	\frac{1}{( \T-\q+1 )} \sum_{t=\q^\prime}^{\T-\q+\q^\prime} w_{it}
	\right)
\\
& =
\frac{1}{\q} \sum_{\q^\prime=1}^{\q}
\left(
\frac{1}{\T} \Big( \sum^{\q^\prime-1}_{t=1} w_{it} + \sum_{t=\T-\q+\q^\prime+1}^{\T} w_{it} \Big)
-
\frac{( \q-1 )}{\T( \T-\q+1 )} \sum_{t=0}^{\T-\q} w_{it+\q^\prime}
\right), 
\end{split}
\end{equation*}
which follows from breaking up the sample average $\bar{w}_i = \nicefrac{1}{\T} \sum_{t=1}^{\T} w_{it}$ into three blocks and re-arranging. Then, by Jensen's inequality and the fact that  $( x+y+z )^2 \leq 3( x^2+y^2+z^2 )$, 
\begin{equation*}
\begin{split}
\mathbb{E}\left( \lvert \bar{w}_i - \mathbb{E}^*( \bar{w}_i^* ) \rvert^{2} \right)
& \leq
\frac{1}{\q} \sum_{\q^\prime=1}^{\q}
\frac{3}{\T^2} \, \mathbb{E}\left( \left\lvert \sum^{\q^\prime-1}_{t=1} w_{it} \right\rvert^{2}  \right)
 +
\frac{1}{\q} \sum_{\q^\prime=1}^{\q}
\frac{3}{\T^2} \, \mathbb{E}\left( \left\lvert \sum_{t=\T-\q+\q^\prime+1}^{\T} w_{it} \right\rvert^{2} \right) 
\\
& +
\frac{1}{\q} \sum_{\q^\prime=1}^{\q}
\frac{3( \q-1 )^2}{\T^2( \T-\q+1 )^2} \, \mathbb{E}\Big( \Big\lvert \sum_{t=0}^{\T-\q} w_{it+\q^\prime} \Big\rvert^{2} \Big).
\end{split}
\end{equation*}
We can apply Part (a) to each of the three terms on the right-hand side. This then yields
$$
\mathbb{E}\left( \lvert \bar{w}_i - \mathbb{E}^*( \bar{w}_i^* ) \rvert^{2} \right)
\leq 
	C \, \frac{3( \q-1 )}{\T^2} \left( 1 + \frac{( \q-1 )}{( \T-\q+1 )} \right)
	=
	C \, \frac{3( \q-1 )}{\T( \T-\q+1 )}
	\leq
	6C  \, \frac{ \q}{\T^{2}},
$$
again using $q \leq \nicefrac{\T}{2}$ in the final step. This proves Part (c) of the lemma and, therefore, completes the proof. 
\end{proof}

\begin{lemma} \label{lemma:cross}
Let Assumption \ref{ass:ass1} hold and take $\q \leq \nicefrac{\T}{2}$. Consider the two random variables
$$
X_{i\varpi}
\coloneqq
\frac{1}{\sqrt{\q}} \sum_{\q^\prime=1}^{\q} ( z_{i\varpi+\q^\prime} - \mathbb{E}^*( \bar{z}_i^* ) ),
\qquad 
Y_{i\varpi}
\coloneqq
\frac{1}{\sqrt{\q}} \sum_{\q^\prime=1}^{\q} ( \varepsilon_{i\varpi+\q^\prime} - \mathbb{E}^*( \bar{\varepsilon}_i^* ) ).
$$
Then, for all $i \neq j$,

\medskip\noindent
(a)
$\mathbb{E}( \mathbb{E}^*( X_{i\varpi}^4 ) ) \leq C$ and $\mathbb{E}( \mathbb{E}^*( Y_{i\varpi}^4 ) ) \leq C$,

\medskip\noindent
(b)
$\mathbb{E}( \lvert \mathbb{E}^*( X_{i\varpi} X_{j\varpi} ) \rvert^2 ) \leq C \, \nicefrac{\q}{\T}$,
$\mathbb{E}( \lvert \mathbb{E}^*( Y_{i\varpi} Y_{j\varpi} ) \rvert^2 ) \leq C \, \nicefrac{\q}{\T}$, and $\mathbb{E}( \lvert \mathbb{E}^*( X_{i\varpi} Y_{j\varpi} ) \rvert^2 ) \leq C \, \nicefrac{\q}{\T}$, 

\medskip\noindent
(c)
$\lvert \mathbb{E}( \mathbb{E}^*( X_{i\varpi} Y_{i\varpi} ) ) - \mathbb{E}( \chi_i ) \rvert \leq C \, ( \nicefrac{1}{\q} + \nicefrac{\q}{\T} )$,

\medskip\noindent
where the finite constant $C$ does not depend on $i$, $j$, $\q$, or $\T$.
\end{lemma}

\begin{proof}
To show Part (a), we first use the fact that $( x - y )^4 \leq 2^3( x^4 + y^4 )$ to get the upper bound
$$
X_{i\varpi}^4
=
\left(\frac{1}{\sqrt{\q}} \sum_{\q^\prime=1}^{\q} z_{i\varpi+\q^\prime} - \sqrt{\q} \, \mathbb{E}^*( \bar{z}_i^* ) \right)^4
\leq
2^3 \left( \frac{1}{\q^2} \Big( \sum_{\q^\prime=1}^{\q} z_{i\varpi+\q^\prime} \Big)^4 + \q^2 \, \mathbb{E}^*( \bar{z}_i^* )^4 \right). 
$$
Here, recalling that the $\varpi_1,\ldots ,\varpi_p$ are independent and identically distributed and using Jensen's inequality gives
$$
\q^2 \, \mathbb{E}^*( \bar{z}_i^* )^4
=
\q^2 \, \left( \frac{1}{\p\q} \sum_{\p^\prime=1}^{\p} \mathbb{E}^*\left( \sum_{\q^\prime=1}^{\q} z_{i\varpi_{\p^\prime}+\q^\prime} \right) \right)^4
\leq
\mathbb{E}^*\left( \frac{1}{\q^2} \left( \sum_{\q^\prime=1}^{\q} z_{i\varpi+\q^\prime} \right)^4 \right) .
$$
Therefore, using Lemma \ref{lemma:Estar}, 
$$
\mathbb{E}^*( X_{i\varpi}^4 )
\leq 
2^4 \, \mathbb{E}^*\left( \frac{1}{\q^2} \left( \sum_{\q^\prime=1}^{\q} z_{i\varpi+\q^\prime} \right)^4 \right) 
=
2^4  \, 
\frac{1}{\q^2} \, \frac{1}{( \T-\q+1 )} \sum_{t=0}^{\T-\q} \left( \sum_{\q^\prime=1}^{\q} z_{it+\q^\prime} \right)^4.
$$
By an application of Part (a) of Lemma \ref{lemma:wsum} (with $w_{it} = z_{it}$, $a = 2$, and $\ell = \q$) we then obtain
$$
\mathbb{E}(\mathbb{E}^*( X_{i\varpi}^4 )) \leq 2^4 \, C .
$$
Exactly the same argument can be used to verify that 
$
\mathbb{E}(\mathbb{E}^*( Y_{i\varpi}^4 )) \leq 2^4 \, C ,
$
and so the proof of Part (a) is complete.

Moving on to Part (b) we focus on establishing that 
$$
\mathbb{E}( \lvert \mathbb{E}^*( X_{i\varpi} Y_{j\varpi} ) \rvert^2 ) \leq C \, \q \T^{-1} ;
$$
the two other statements follow in the same way. We begin by expanding the product by writing
$$
X_{i\varpi} Y_{j\varpi}
 =
\frac{1}{\q} \sum_{\q^\prime=1}^{\q} 
z_{i\varpi+\q^\prime} \sum_{\q^{\prime\prime}=1}^{\q} \varepsilon_{j\varpi+\q^{\prime\prime}}
 - 
\mathbb{E}^*( \bar{\varepsilon}_j^* ) \sum_{\q^\prime=1}^{\q} z_{i\varpi+\q^\prime}
-
\mathbb{E}^*( \bar{z}_i^* ) \sum_{\q^{\prime\prime}=1}^{\q} \varepsilon_{j\varpi+\q^{\prime\prime}} 
+
\q \, \mathbb{E}^*( \bar{z}_i^* ) \, \mathbb{E}^*( \bar{\varepsilon}_j^* ).
$$
Observing that $\mathbb{E}^*( \sum_{\q^\prime=1}^{\q} z_{i\varpi+\q^\prime} ) = \q \, \mathbb{E}^*( \bar{z}_i^* )$ and $\mathbb{E}^*( \sum_{\q^{\prime\prime}=1}^{\q} \varepsilon_{j\varpi+\q^{\prime\prime}} ) = \q \, \mathbb{E}^*( \bar{\varepsilon}_j^* )$, and using Lemma \ref{lemma:Estar},
$$
	\mathbb{E}^*( X_{i\varpi} Y_{j\varpi} )
	=\
	\frac{1}{\q( \T-\q+1 )} \sum_{t=0}^{\T-\q}  \sum_{\q^\prime=1}^{\q} z_{it+\q^\prime} \sum_{\q^{\prime\prime}=1}^{\q} \varepsilon_{jt+\q^{\prime\prime}}
	-
	\q \, \mathbb{E}^*( \bar{z}_i^* ) \, \mathbb{E}^*( \bar{\varepsilon}_j^* ). 
$$
Using the fact that $( x - y )^2 \leq 2x^2 + 2y^2$ and exploiting that the data are independent across $i$ we can bound $\mathbb{E}\left( \lvert \mathbb{E}^*( X_{i\varpi} Y_{j\varpi} ) \rvert^2 \right)$ from above by  
$$
2 \, \mathbb{E}\left( \left( \frac{1}{\q( \T-\q+1 )} \sum_{t=0}^{\T-\q}  \sum_{\q^\prime=1}^{\q} z_{it+\q^\prime} \sum_{\q^{\prime\prime}=1}^{\q} \varepsilon_{jt+\q^{\prime\prime}} \right)^2 \right) 
	+\ 2 \q^2 \, \mathbb{E}\left( \mathbb{E}^*( \bar{z}_i^* )^2 \right) \, \mathbb{E}\left( \mathbb{E}^*( \bar{\varepsilon}_j^* )^2 \right).
$$
Consider first the second term in this bound. By an application of Part (b) of Lemma \ref{lemma:wsum} (with $a = 1$, $w_{it} = z_{it}$, and $w_{it} = \varepsilon_{it}$) we have
$$
\mathbb{E}\left( \mathbb{E}^*( \bar{z}_i^* )^2 \right) \leq \frac{C}{\T},  
\qquad 
\mathbb{E}\left( \mathbb{E}^*( \bar{\varepsilon}_j^* )^2 \right) \leq \frac{C}{\T},
$$
from which 
$
2 \q^2 \, 
\mathbb{E}\left( \mathbb{E}^*( \bar{z}_i^* )^2 \right) \, \mathbb{E}\left( \mathbb{E}^*( \bar{\varepsilon}_j^* )^2 \right) \leq 2 C^2 \, (\nicefrac{ \q}{\T})^2 \leq C^2 \, \nicefrac{ \q}{\T}
$
then follows. For the first term in the bound, in turn, expanding the square, and using independence over $i$ once again,
\begin{align*}
	\mathbb{E}\Big( \Big( & \frac{1}{\q( \T-\q+1 )} \sum_{t=0}^{\T-\q}  \sum_{\q^\prime=1}^{\q} z_{it+\q^\prime} \sum_{\q^{\prime\prime}=1}^{\q} \varepsilon_{jt+\q^{\prime\prime}} \Big)^2 \Big) \notag \\
	= &
	\frac{1}{\q^2( \T-\q+1 )^2} \sum_{t=0}^{\T-\q} \sum_{t^\prime=0}^{\T-\q} 
	\mathbb{E}\Big( \sum_{\q^\prime=1}^{\q} z_{it+\q^\prime} \sum_{\q^{\prime\prime}=1}^{\q} z_{it^\prime+\q^{\prime\prime}} \sum_{\q^\prime=1}^{\q} \varepsilon_{jt+\q^\prime} \sum_{\q^{\prime\prime}=1}^{\q} \varepsilon_{jt^\prime+\q^{\prime\prime}} \Big) \notag \\
	= &
	\frac{1}{\q^2( \T-\q+1 )^2} \sum_{t=0}^{\T-\q} \sum_{t^\prime=0}^{\T-\q} 
	\mathbb{E}\Big( \sum_{\q^\prime=1}^{\q} z_{it+\q^\prime} \sum_{\q^{\prime\prime}=1}^{\q} z_{it^\prime+\q^{\prime\prime}} \Big) \,
	\mathbb{E}\Big( \sum_{\q^\prime=1}^{\q} \varepsilon_{jt+\q^\prime} \sum_{\q^{\prime\prime}=1}^{\q} \varepsilon_{jt^\prime+\q^{\prime\prime}} \Big) \notag \\
	\leq &
	\frac{1}{\q^2( \T-\q+1 )^2} \sum_{t=0}^{\T-\q} 
	\Big( \sum_{t^\prime=0}^{\T-\q} \Big\lvert \mathbb{E}\Big( \sum_{\q^\prime=1}^{\q} z_{it+\q^\prime} \sum_{\q^{\prime\prime}=1}^{\q} z_{it^\prime+\q^{\prime\prime}} \Big) \Big\rvert \Big) 
\\
 \times &	\max_{0 \leq t, t^\prime \leq \T-\q} \Big\lvert \mathbb{E}\Big( \sum_{\q^\prime=1}^{\q} \varepsilon_{jt+\q^\prime} \sum_{\q^{\prime\prime}=1}^{\q} \varepsilon_{jt^\prime+\q^{\prime\prime}} \Big) \Big\rvert .
\end{align*}
Using the Cauchy-Schwarz inequality and Lemma \ref{lemma:wsum} Part (a) (with $w_{it} = \varepsilon_{it}$, $a = 1$, and $\ell = \q$) gives
\begin{equation*}
\begin{split}
\max_{0 \leq t, t^\prime \leq \T-\q} &
\left\lvert \mathbb{E}\left( \sum_{\q^\prime=1}^{\q} \varepsilon_{jt+\q^\prime} \sum_{\q^{\prime\prime}=1}^{\q} \varepsilon_{jt^\prime+\q^{\prime\prime}} \right) 
\right\rvert
\\
& \leq
\max_{0 \leq t, t^\prime \leq \T-\q} 
\sqrt{\mathbb{E}\left( \left( \sum_{\q^\prime=1}^{\q} \varepsilon_{jt+\q^\prime} \right)^2 \right)} \, \sqrt{
\mathbb{E}\left( \left( \sum_{\q^{\prime\prime}=1}^{\q} \varepsilon_{jt^\prime+\q^{\prime\prime}} \right)^2 \right) }
\leq C \, \q .
\end{split}
\end{equation*}
Also,
$$
\sum_{t^\prime=0}^{\T-\q} 
\left\lvert 
\mathbb{E}\left( \sum_{\q^\prime=1}^{\q} z_{it+\q^\prime} \sum_{\q^{\prime\prime}=1}^{\q} z_{it^\prime+\q^{\prime\prime}} \right) 
\right\rvert
\leq 
\sum_{\q^\prime=1}^{\q} \sum_{\q^{\prime\prime}=1}^{\q} \sum_{t^\prime=0}^{\T-\q} \lvert \mathbb{E}( z_{it+\q^\prime} z_{it^\prime+\q^{\prime\prime}} ) \rvert
\leq 
\q^2 \,  \sum_{h=-\infty}^{+\infty} \sup_{i,t} \lvert \mathbb{E}( z_{it} z_{it+h} ) \rvert .
$$
Here, by Davydov's inequality (see, e.g., \citealt[Corollary A.2]{HallHeyde1980}), we have the bound
$$
\sup_{i,t} \lvert \mathbb{E}( z_{it} z_{it+h} ) 
\leq
8 \, \bar{a}( \lvert h \rvert )^{1-\nicefrac{2}{r}} \, \sup_{i,t} \mathbb{E}( \lvert z_{it} \rvert^r )^{\nicefrac{2}{r}} .
$$
As $\bar{a}( h ) = O( h^{-s} )$ with $s( 1-\nicefrac{2}{r} ) > 1$, 
$
\sum_{h=-\infty}^{+\infty} \bar{a}( \lvert h \rvert )^{1-\nicefrac{2}{r}}
\leq 
2 C \sum_{h=1}^{\infty} h^{-s(1-\nicefrac{2}{r})}
< \infty,
$
and so
$$
\sum_{t^\prime=0}^{\T-\q} 
\left\lvert 
\mathbb{E}\left( \sum_{\q^\prime=1}^{\q} z_{it+\q^\prime} \sum_{\q^{\prime\prime}=1}^{\q} z_{it^\prime+\q^{\prime\prime}} \right) 
\right\rvert \leq C \, \q^2.
$$
Combining results yields
$$
\mathbb{E}\left( \lvert \mathbb{E}^*( X_{i\varpi} Y_{j\varpi} ) \rvert^2 \right)
\leq 
2 C^2 \, \frac{ \q}{( \T-\q+1 )} + C^2  \, \frac{\q}{\T} 	\leq 5 C^2 \, 	\frac{ \q}{\T},
$$ 
as claimed in Part (b) of the lemma.

Finally, for Part (c) we again begin by expanding and taking expectations to arrive at
$$
\mathbb{E}\left( \mathbb{E}^*( X_{i\varpi}Y_{i\varpi} )\right)
=
\frac{1}{\q ( \T-\q+1 )}\sum_{t=0}^{\T-\q}
\sum_{\q^\prime=1}^{\q} \sum_{\q^{\prime\prime}=1}^{\q}  
\mathbb{E}
\left( 
z_{it+\q^\prime} 
\varepsilon_{it+\q^{\prime\prime}}
\right) 
-
\q \, \mathbb{E}\left( \mathbb{E}^*( \bar{z}_i^* )\,\mathbb{E}^*( \bar{\varepsilon}_i^* )\right).
$$
Here,
\begin{equation*}
\begin{split}
\frac{1}{\q} \,
\sum_{\q^\prime=1}^{\q} \sum_{\q^{\prime\prime}=1}^{\q}  
\mathbb{E}
\left( 
z_{it+\q^\prime} 
\varepsilon_{it+\q^{\prime\prime}}
\right)
& = 
\frac{1}{q} \sum_{\lvert h \rvert < q} (q- \lvert h \rvert) \, \mathbb{E}(z_{it} \varepsilon_{it+h})
=
 \sum_{\lvert h \rvert < q} \left(1- \frac{\lvert h \rvert}{\q} \right) \, \mathbb{E}(z_{it} \varepsilon_{it+h}).
\end{split}
\end{equation*}
This follows because, by stationarity, $\mathbb{E}( z_{it+\q^\prime} \varepsilon_{it+\q^{\prime\prime}} ) = \mathbb{E}( z_{it} \varepsilon_{it+h} )$ depends on $( \q^\prime, \q^{\prime\prime} )$ only through the difference $h = \q^{\prime\prime} - \q^\prime$ and there are $\q - \lvert h \rvert$ pairs $( \q^\prime, \q^{\prime\prime} )$ with $\q^{\prime\prime} - \q^\prime = h$ for each $\lvert h \rvert < \q$.
Also, as
$$
\chi_i 
= \frac{1}{\T} \sum_{t=1}^{\T} \sum_{t^\prime=1}^{\T} z_{it} \varepsilon_{it^\prime},
$$
we have, in the same way, that
$$
\mathbb{E}(\chi_i)
=
\sum_{\lvert h \rvert < \T} \left(1 - \frac{\lvert h \rvert}{\T} \right) \, \mathbb{E}(z_{it} \varepsilon_{it+h}).
$$
We can thus write
\begin{equation*}
\begin{split}
\mathbb{E}\left( \mathbb{E}^*( X_{i\varpi} Y_{i\varpi} ) \right) - \mathbb{E}( \chi_i )
& = \hspace{.20cm}
\sum_{\lvert h \rvert < \q} \left( \frac{\lvert h \rvert}{\T} - \frac{\lvert h \rvert}{\q} \right) \mathbb{E}( z_{it} \varepsilon_{it+h} )
\\
& -
\sum_{\q \leq \lvert h \rvert < \T} \left( 1 - \frac{\lvert h \rvert}{\T} \right) \mathbb{E}( z_{it} \varepsilon_{it+h} )
 - 
\q \, \mathbb{E}\left( \mathbb{E}^*( \bar{z}_i^* ) \, \mathbb{E}^*( \bar{\varepsilon}_i^* ) \right) .
\end{split}
\end{equation*}
For the last right-hand side term we can use the Cauchy-Schwarz inequality followed by an application of Lemma \ref{lemma:wsum} Part (b) (with $a = 1$, $w_{it} = z_{it}$, and $w_{it} = \varepsilon_{it}$) to see that it is bounded by $C \, \nicefrac{\q}{\T}$. For the remaining two terms, note that, when $\lvert h \rvert < \q$ we have $\lvert \nicefrac{\lvert h \rvert}{\T} - \nicefrac{\lvert h \rvert}{\q} \rvert \leq \nicefrac{\lvert h \rvert}{\q}$ whereas, when $\lvert h \rvert \geq \q$, we have $1 - \nicefrac{\lvert h \rvert}{\T} \leq 1 \leq \nicefrac{\lvert h \rvert}{\q}$. Consequently, we can collect terms to obtain
$$
\left\lvert \mathbb{E}\left( \mathbb{E}^*( X_{i\varpi} Y_{i\varpi} ) \right) - \mathbb{E}( \chi_i ) \right\rvert
\leq 
\frac{1}{\q} \sum_{h=-\infty}^{+\infty} \lvert h \rvert \, \lvert \mathbb{E}( z_{it} \varepsilon_{it+h} ) \rvert + C\, \frac{\q}{\T}.
$$
By another application  of Davydov's inequality, 
$$
\lvert \mathbb{E}( z_{it} \varepsilon_{it+h} ) \rvert
\leq 
8 \,  \bar{a}( \lvert h \rvert )^{1-\nicefrac{2}{r}} \, \sup_{i,t} \mathbb{E}( \lvert z_{it} \rvert^r )^{\nicefrac{1}{r}} \, \sup_{i,t} \mathbb{E}( \lvert \varepsilon_{it} \rvert^r )^{\nicefrac{1}{r}},
$$
and, as $\bar{a}( h ) = O( h^{-s} )$ with $s( 1-\nicefrac{2}{r} ) > 2$, 
$
\sum_{h=-\infty}^{+\infty} \lvert h \rvert \, \bar{a}( \lvert h \rvert )^{1-\nicefrac{2}{r}}
\leq 
2 C \sum_{h=1}^{\infty} h^{1-s(1-\nicefrac{2}{r})} < \infty, 
$
so that
$
\nicefrac{1}{\q} \sum_{h=-\infty}^{+\infty} \lvert h \rvert \, \lvert \mathbb{E}( z_{it} \varepsilon_{it+h} ) \rvert
\leq C \, \nicefrac{1}{\q}.
$
Thus, 
$$
\left\lvert \mathbb{E}\left( \mathbb{E}^*( X_{i\varpi} Y_{i\varpi} ) \right) - \mathbb{E}( \chi_i ) \right\rvert
\leq 
C \, (\q^{-1} + \q \T^{-1}), 
$$
which is Part (c) of the lemma. The proof is then complete.
\end{proof}

\paragraph{Proof of Theorem \ref{thm:thm2}.}
\medskip
From the definition of $\hat{\beta}^*$,
\begin{align}
	\sqrt{\N\T} ( \hat{\beta}^* - \hat{\beta} )
	=
	( \hat{\varSigma}_{\N,\T}^* )^{-1}
	\left(
	\frac{1}{\sqrt{\N\T}} \sum_{i=1}^{\N} \sum_{t=1}^{\T} ( x_{it}^* - \bar{x}^*_i ) \, \hat{\varepsilon}_{it}^*
	\right), \label{eq:betastar}
\end{align}
where
\begin{align}
	\hat{\varSigma}_{\N,\T}^*
	= 
	\frac{1}{\N\T} \sum_{i=1}^{\N} \sum_{t=1}^{\T} ( x_{it}^* - \bar{x}_i^* )^2 \notag
\end{align}
and
$
\hat{\varepsilon}_{i \, (\p^\prime-1)\q+\q^\prime}^* \coloneqq \hat{\varepsilon}_{i\, \varpi_{\p^\prime}+\q^\prime},
$
for $1\leq \p^\prime \leq \p$ and $1\leq \q^\prime \leq \q$,
are the resampled residuals associated with the within-group estimator computed from the original data. Moreover, $\hat{\varepsilon}_{it} \coloneqq ( y_{it} - \bar{y}_i ) - ( x_{it}-\bar{x}_i )\hat{\beta}$. As in the proof of Theorem \ref{thm:thm1} we will proceed in multiple steps.

We first show that $\lvert \hat{\varSigma}_{\N,\T}^* - \hat{\varSigma}_{\N,\T} \rvert = o_{P^*}( 1 )$, that is,
\begin{align}
	\mathbb{P}( \mathbb{P}^*( \lvert \hat{\varSigma}_{\N,\T}^* - \hat{\varSigma}_{\N,\T} \rvert > \epsilon^* ) > \epsilon )
	=
	o( 1 ) \notag
\end{align}
for all $\epsilon^*>0$ and $\epsilon>0$. Recall that $z_{it} = x_{it} - \mathbb{E}( x_{it} )$, so, by stationarity, $\bar{x}_i^* - \mathbb{E}( \bar{x}_i ) = \bar{z}_i^*$ and $\bar{x}_i - \mathbb{E}( \bar{x}_i ) = \bar{z}_i$. Adding and subtracting $\mathbb{E}( x_{it}^2 )$ and re-arranging yields the decomposition
\begin{equation*}
\begin{split}
\hat{\varSigma}_{\N,\T}^* - \hat{\varSigma}_{\N,\T}
& = 
\frac{1}{\N\T}\sum_{i=1}^{\N}\sum_{t=1}^{\T} \left( {x_{it}^*}^{2} - \mathbb{E}( x_{it}^2 ) \right)
 -
\frac{1}{\N\T}\sum_{i=1}^{\N}\sum_{t=1}^{\T} \left( x_{it}^{2} - \mathbb{E}( x_{it}^2 ) \right)
\\
& -
 \frac{1}{\N}\sum_{i=1}^{\N} {\bar{z}_i^*}^2 
 -
 \frac{2}{\N}\sum_{i=1}^{\N} \mathbb{E}( \bar{x}_i ) \, ( \bar{z}_i^* - \bar{z}_i ) 
  +
 \frac{1}{\N}\sum_{i=1}^{\N} \bar{z}_i^2,
\end{split}
\end{equation*}
and we show that each of the five right-hand side terms is $o_{P^*}(1)$, in turn. For the first term we have, using the triangle inequality, the Cauchy-Schwarz inequality, and  Lemma \ref{lemma:Estar} and Lemma \ref{lemma:wsum} Part (a) (with $a = 1$, $\ell = \q$, and $w_{it} = x_{it}^2 - \mathbb{E}( x_{it}^2 )$),
\begin{align}
	&\mathbb{E}\Big( \mathbb{E}^*\Big( \Big\lvert \frac{1}{\N\T}\sum_{i=1}^{\N}\sum_{t=1}^{\T} \big( {x_{it}^*}^{2} - \mathbb{E}( x_{it}^2 ) \big) \Big\rvert \Big) \Big) \notag \\
	&\quad\leq
	\frac{1}{\N} \sum_{i=1}^{\N} \frac{1}{\p} \sum_{\p^\prime=1}^{\p}
	\mathbb{E}\Big( \mathbb{E}^*\Big( \Big\lvert \frac{1}{\q} \sum_{\q^\prime=1}^{\q} \big( x_{i\varpi_{\p^\prime}+\q^\prime}^{2} - \mathbb{E}( x_{it}^{2} ) \big) \Big\rvert \Big) \Big) \notag \\
	&\quad=
	\frac{1}{\N} \sum_{i=1}^{\N} \frac{1}{( \T-\q+1 )} \sum_{t=0}^{\T-\q}
	\mathbb{E}\Big( \Big\lvert \frac{1}{\q} \sum_{\q^\prime=1}^{\q} \big( x_{it+\q^\prime}^{2} - \mathbb{E}( x_{it+\q^\prime}^{2} ) \big) \Big\rvert \Big) \notag \\
	&\quad\leq
	\sup_{i} \, \sup_{0\leq t \leq \T-\q}
	\Big( \mathbb{E}\Big( \Big\lvert \frac{1}{\q} \sum_{\q^\prime=1}^{\q} \big( x_{it+\q^\prime}^{2} - \mathbb{E}( x_{it+\q^\prime}^{2} ) \big) \Big\rvert^2 \Big) \Big)^{\nicefrac{1}{2}} 
	\leq
	C^{\nicefrac{1}{2}} \, \q^{-\nicefrac{1}{2}} . \notag
\end{align}
The first term is then $o_{P^*}(1)$ by an application of Markov's inequality. We follow the same strategy for each of the remaining terms. For the second term, using the same arguments, 
\begin{equation*}
\begin{split}
\mathbb{E}\left( \mathbb{E}^*\left( \left\lvert \frac{1}{\N\T}\sum_{i=1}^{\N}\sum_{t=1}^{\T} \left( x_{it}^{2} - \mathbb{E}( x_{it}^2 ) \right) \right\rvert \right) \right) 
\leq
\sup_{i}
\left( \mathbb{E}\left( \left\lvert \frac{1}{\T} \sum_{t=1}^{\T} \left( x_{it}^{2} - \mathbb{E}( x_{it}^2 ) \right) \right\rvert^2 \right) \right)^{\nicefrac{1}{2}} 
\end{split}
\end{equation*}
 is $O(\nicefrac{1}{\sqrt{\T}})$ by  an application of Lemma \ref{lemma:wsum} Part (a) (with $a = 1$, $\ell = \T$, and $w_{it} = x_{it}^2 - \mathbb{E}( x_{it}^2 )$). For the third term, using both parts of Lemma \ref{lemma:Estar} gives 
 \begin{equation*}
 \begin{split}
 \mathbb{E}\left( \mathbb{E}^*\left( \frac{1}{\N}\sum_{i=1}^{\N} {\bar{z}_i^*}^2 \right) \right)
 & =
 \frac{1}{\N} \sum_{i=1}^{\N}
\frac{1}{\p} \, \mathbb{E}\left( \mathbb{E}^*\left( \left\lvert \frac{1}{\q} \sum_{\q^\prime=1}^{\q} z_{i\varpi_1+\q^\prime} - \mathbb{E}^*( \bar{z}_i^* ) \right\rvert^2 \right) \right)+\mathbb{E}( \lvert \mathbb{E}^*( \bar{z}_i^* ) \rvert^2 )
\\
& \leq \
\sup_{i}
\left(
 \frac{1}{\p\q^2 ( \T-\q+1 )} \sum_{t=0}^{\T-\q} \mathbb{E}\left( \left\lvert \sum_{\q^\prime=1}^{\q} z_{it+\q^\prime} \right\rvert^2 \right)
+
\mathbb{E}( \lvert \mathbb{E}^*( \bar{z}_i^* ) \rvert^2 )
\right) .
 \end{split}
 \end{equation*}
An application of Lemma \ref{lemma:wsum} Part (a) (with $a = 1$, $\ell = \q$, and $w_{it} = z_{it}$) and  of Lemma \ref{lemma:wsum} Part (b) (with $a = 1$ and $w_{it} = z_{it}$) then reveals that 
$$
 \mathbb{E}\left( \mathbb{E}^*\left( \frac{1}{\N}\sum_{i=1}^{\N} {\bar{z}_i^*}^2 \right) \right)
 =
 O(\T^{-1})
 $$
on recalling that $\T = \p\q$. For the fourth term, 
$$
\mathbb{E}\left( \mathbb{E}^*\left( \left\lvert \frac{2}{\N}\sum_{i=1}^{\N} \mathbb{E}( \bar{x}_i ) \, ( \bar{z}_i^* - \bar{z}_i ) \right\rvert \right) \right)
\leq
2 
\sup_{i} 
\lvert \mathbb{E}( \bar{x}_i ) \rvert 
\,
\sup_{i} 
\sqrt{ \mathbb{E}\left( \mathbb{E}^*( ( \bar{z}_i^* - \bar{z}_i )^2 ) \right) },
$$
where we used the Cauchy-Schwarz inequality. Here, $\sup_{i}  \lvert \mathbb{E}( \bar{x}_i ) \rvert$ is uniformly bounded by assumption. Furthermore, as $(x-y)^2 \leq 2 (x^2 + y^2)$, 
$$
\mathbb{E}\left( \mathbb{E}^*( ( \bar{z}_i^* - \bar{z}_i )^2 ) \right) 
\leq
2 \, \left( \mathbb{E}\big( \mathbb{E}^*( ( \bar{z}_i^* )^2 ) \big) +  \mathbb{E}( \bar{z}_i^2 ) \right).
$$
Above we had already shown that
$\mathbb{E}( \mathbb{E}^*( ( \bar{z}_i^* )^2 ) ) \leq 2C \, \nicefrac{1}{\T}$ whereas, by an application of Lemma \ref{lemma:wsum} Part (a) (with $w_{it} = z_{it}$, $a = 1$, $\ell = \T$, and $t = 0$), it likewise follows that
$
	\mathbb{E}( \bar{z}_i^2 )
	=
	\nicefrac{1}{\T^2} \, \mathbb{E}( ( \sum_{l=1}^{\T} z_{il} )^2 )
	\leq
	C \, \nicefrac{1}{\T}. 
$
Hence, 
$$
\mathbb{E}\left( \mathbb{E}^*\left( \left\lvert \frac{2}{\N}\sum_{i=1}^{\N} \mathbb{E}( \bar{x}_i ) \, ( \bar{z}_i^* - \bar{z}_i ) \right\rvert \right) \right) = O(\T^{-\nicefrac{1}{2}})
$$
follows. The fifth and final term follows easily, as
$$
\mathbb{E}\left( \mathbb{E}^* \left( \frac{1}{\N}\sum_{i=1}^{\N} \bar{z}_i^2 \right) \right) = \frac{1}{\N}\sum_{i=1}^{\N} \mathbb{E}( \bar{z}_i^2 ) 
\leq \frac{C}{\T} .
$$
Therefore, we have shown that $\lvert \hat{\varSigma}^*_{\N,\T} - \hat{\varSigma}_{\N,\T} \rvert = o_{P^*}( 1 )$, which completes our analysis of the denominator in  \eqref{eq:betastar}.

We now turn to the numerator in \eqref{eq:betastar},
$$
\frac{1}{\sqrt{\N\T}} \sum_{i=1}^{\N} \sum_{t=1}^{\T} ( x_{it}^* - \bar{x}^*_i ) \, \hat{\varepsilon}_{it}^*. 
$$
First, by the first-order condition of the within-group estimator, 
$$
\frac{1}{\sqrt{\N\T}} \sum_{i=1}^{\N} \sum_{t=1}^{\T} ( x_{it} - \bar{x}_i ) \, \hat{\varepsilon}_{it} = 0, 
$$
where we recall that $\nicefrac{1}{\T} \sum_{t=1}^{\T}  \hat{\varepsilon}_{it} = 0$ for all $1\leq i \leq \N$ by definition of the within-group estimator. Therefore,
$$
\frac{1}{\sqrt{\N\T}} \sum_{i=1}^{\N} \sum_{t=1}^{\T} ( x_{it}^* - \bar{x}^*_i ) \hat{\varepsilon}_{it}^*
=
\frac{1}{\sqrt{\N\T}} \sum_{i=1}^{\N} \sum_{t=1}^{\T}
\left(
( x_{it}^* - \bar{x}^*_i ) \, \hat{\varepsilon}_{it}^*
-
( x_{it} - \bar{x}_i ) \, \hat{\varepsilon}_{it}
\right). 
$$
Next,
$
\hat{\varepsilon}_{it}
=
( y_{it} - \bar{y}_i ) - ( x_{it} - \bar{x}_i ) \hat{\beta}
=
- ( x_{it} - \bar{x}_i ) ( \hat{\beta} - \beta )
+
( \varepsilon_{it} - \bar{\varepsilon}_i )
$
by definition of the within-group residuals. Furthermore, by definition of the moving block bootstrap, equally,
$$
\hat{\varepsilon}_{it}^* =	- ( x_{it}^* - \bar{x}_i ) ( \hat{\beta} - \beta )+ ( \varepsilon_{it}^* - \bar{\varepsilon}_i ), 
$$
where
$
{\varepsilon}_{i \, (\p^\prime-1)\q+\q^\prime}^* \coloneqq {\varepsilon}_{i\, \varpi_{\p^\prime}+\q^\prime}
$
for $1\leq \p^\prime \leq \p$ and $1\leq \q^\prime \leq \q$. Because $\sum_{t=1}^{\T} ( x_{it}^* - \bar{x}_i^* ) = 0$, the means $\bar{x}_i$ and $\bar{\varepsilon}_i$ appearing in this final expression can be replaced by $\bar{x}_i^*$ and $\bar{\varepsilon}_i^*$ once we multiply through by $x_{it}^* - \bar{x}_i^*$ and sum over $t$. We thus obtain
\begin{equation}
	\frac{1}{\sqrt{\N\T}} \sum_{i=1}^{\N} \sum_{t=1}^{\T} ( x_{it}^* - \bar{x}^*_i ) \, \hat{\varepsilon}_{it}^*
	=
	\frac{1}{\sqrt{\N\T}} \sum_{i=1}^{\N} \sum_{t=1}^{\T}
	\left(
	( x_{it}^* - \bar{x}^*_i ) \, {\varepsilon}_{it}^*
	-
	( x_{it} - \bar{x}_i ) \, {\varepsilon}_{it}
	\right)
	+
	o_{P^*}( 1 ), \label{eq:bstrapscore}
\end{equation}
where the $o_{P^*}( 1 )$ term is equal to
$$
-
\left(
\frac{1}{\N\T} \sum_{i=1}^{\N} \sum_{t=1}^{\T} ( x_{it}^* - \bar{x}_i^* )^2 - ( x_{it} - \bar{x}_i )^2
\right) \,  	
\sqrt{\N\T} ( \hat{\beta} - \beta ); 
$$
the term in parentheses is equal to $\hat{\varSigma}_{\N,\T}^*-\hat{\varSigma}_{\N,\T}$, which has already been shown to be $o_{P^*}( 1 )$, and $\sqrt{\N\T} ( \hat{\beta}-\beta ) = O_P( 1 )$ from Theorem \ref{thm:thm1}. It thus remains only to analyse the leading term in \eqref{eq:bstrapscore}. Re-centering both $x_{it}^*$ and $x_{it}$ around $\mathbb{E}( x_{it} )$ and re-arranging terms allows us to write this leading term as
\begin{align}
	\frac{1}{\sqrt{\N\T}} \sum_{i=1}^{\N} \sum_{t=1}^{\T}
	\left(
	z_{it}^* {\varepsilon}_{it}^*
	-
	z_{it} {\varepsilon}_{it}
	\right)
	-
	\left( \sqrt{\frac{\N}{\T}} \right) \, \frac{1}{\N} \sum_{i=1}^{\N} ( \chi_i^* - \chi_i ), \label{eq:bexpansion}
\end{align}
for 
$
\chi_i =
\nicefrac{1}{\T} \sum_{t_1=1}^{\T} \sum_{t_2=1}^{\T} z_{it_1} \varepsilon_{it_2},
$
as previously defined in the proof of Theorem \ref{thm:thm1}, and
\begin{align}
	\chi_i^*
	\coloneqq
	\left( \frac{1}{\sqrt{\T}} \sum_{t=1}^{\T} z_{it}^* \right) \left( \frac{1}{\sqrt{\T}} \sum_{t=1}^{\T} \varepsilon_{it}^* \right)
	=
	\frac{1}{\T} \sum_{t_1=1}^{\T} \sum_{t_2=1}^{\T} z_{it_1}^* \varepsilon_{it_2}^* \notag
\end{align}
its natural bootstrap counterpart. The first term in \eqref{eq:bexpansion} will follow a suitable central limit theorem, in that
$$
\frac{1}{\sqrt{\N\T}} \sum_{i=1}^{\N} \sum_{t=1}^{\T} 
\varOmega_{\N,\T}^{-\nicefrac{1}{2}} \, 
( z_{it}^* \varepsilon_{it}^* - z_{it} \varepsilon_{it} )
\overset{L^*}{\rightarrow} N(0, I) ,
$$
with $L^*$ denoting convergence in law, conditional on the data. The second term, in contrast, will generate the desired bias, in that
$$
- \frac{1}{\N} \sum_{i=1}^{\N} ( \chi_i^* - \chi_i )
= 	b_{\N,\T} + o_{P^*}( 1 ).
$$
To complete the proof it remains to  show each of these claims. We start with the bias term.

To analyze the bias it is useful to introduce
$$
\tilde{X}_{i\varpi}\coloneqq \frac{1}{\sqrt{\q}} \sum_{\q^\prime=1}^{\q}  z_{i\varpi+\q^\prime},
\qquad
\tilde{Y}_{i\varpi}\coloneqq \frac{1}{\sqrt{\q}} \sum_{\q^\prime=1}^{\q}  \varepsilon_{i\varpi+\q^\prime}.
$$
Then
$
\nicefrac{1}{\sqrt{\T}} \sum_{t=1}^{\T} z_{it}^*
=
\nicefrac{1}{\sqrt{p}} \sum_{\p^\prime=1}^\p \tilde{X}_{i\varpi_{\p^\prime}}
$
and
$
\nicefrac{1}{\sqrt{\T}} \sum_{t=1}^{\T} \varepsilon_{it}^*
=
\nicefrac{1}{\sqrt{p}} \sum_{\p^\prime=1}^\p \tilde{Y}_{i\varpi_{\p^\prime}}
$
and the bias is
$$
\chi_i^* 
= 
\frac{1}{p} \sum_{\p_1^\prime=1}^\p \sum_{\p_2^{\prime}=1}^\p \tilde{X}_{i\varpi_{\p_1^\prime}} \tilde{Y}_{i\varpi_{\p_2^{\prime}}}
=
\frac{1}{p} \sum_{\p^\prime=1}^\p  \tilde{X}_{i\varpi_{\p^\prime}} \tilde{Y}_{i\varpi_{\p^{\prime}}}
+
\frac{1}{p} \sum_{\p_1^\prime=1}^\p \sum_{\p_2^{\prime}\neq \p_1^\prime}^\p \tilde{X}_{i\varpi_{\p_1^\prime}} \tilde{Y}_{i\varpi_{\p_2^{\prime}}}.
$$
The first right-hand side term is close to a conventional bootstrap covariance estimator in the sense of \citet{GotzeKunsch1996}. Here, such a covariance estimator would take the form
$$
\hat{\gamma}_i^* \coloneqq 
\frac{1}{\p} \sum_{\p^\prime=1}^\p
(\tilde{X}_{i\varpi_{\p^\prime}} - \bar{\tilde{X}}_{i})
(\tilde{Y}_{i\varpi_{\p^\prime}} - \bar{\tilde{Y}}_{i}),
$$
where $\bar{\tilde{X}}_i \coloneqq \nicefrac{1}{\p} \sum_{\p^\prime=1}^\p \tilde{X}_{i\varpi_{\p^\prime}}$ and $\bar{\tilde{Y}}_i \coloneqq \nicefrac{1}{\p} \sum_{\p^\prime=1}^\p \tilde{Y}_{i\varpi_{\p^\prime}}$. We can then write
$$
\frac{1}{p} \sum_{\p^\prime=1}^\p  \tilde{X}_{i\varpi_{\p^\prime}} \tilde{Y}_{i\varpi_{\p^{\prime}}}
=
\frac{1}{\p} \sum_{\p^\prime=1}^\p
(\tilde{X}_{i\varpi_{\p^\prime}} - \bar{\tilde{X}}_{i})
(\tilde{Y}_{i\varpi_{\p^\prime}} - \bar{\tilde{Y}}_{i})
+
\bar{\tilde{X}}_i \bar{\tilde{Y}}_i 
=
\hat{\gamma}_i^* +
\bar{\tilde{X}}_i \bar{\tilde{Y}}_i .
$$
The second right-hand side term, in turn, factors as
\begin{equation*}
\begin{split}
\frac{1}{\p} \sum_{\p_1^\prime=1}^{\p} \sum_{\p_2^\prime \neq \p_1^\prime} \tilde{X}_{i\varpi_{\p_1^\prime}} \tilde{Y}_{i\varpi_{\p_2^\prime}}
& =
\frac{1}{\p} \sum_{\p_1^\prime=1}^{\p} \sum_{\p_2^\prime \neq \p_1^\prime} {X}_{i\varpi_{\p_1^\prime}} {Y}_{i\varpi_{\p_2^\prime}}
\\
& +
\frac{( \p-1 )}{\p} \sum_{\p^\prime=1}^{\p} X_{i\varpi_{\p^\prime}} \, \mathbb{E}^*( \tilde{Y}_{i\varpi} ) 
+
\frac{( \p-1 )}{\p} \sum_{\p^\prime=1}^{\p} Y_{i\varpi_{\p^\prime}} \, \mathbb{E}^*( \tilde{X}_{i\varpi} )
\\
& +
(\p-1) \, \mathbb{E}^*(\tilde{X}_{i\varpi}) \, \mathbb{E}^*(\tilde{Y}_{i\varpi}) ,
\end{split}
\end{equation*}
where $X_{i\varpi} \coloneqq \tilde{X}_{i\varpi} - \mathbb{E}^*(\tilde{X}_{i\varpi})$ and $Y_{i\varpi} \coloneqq \tilde{Y}_{i\varpi} - \mathbb{E}^*(\tilde{Y}_{i\varpi})$. Combining all terms then yields
\begin{equation} \label{eq:biasde}
\begin{split}
(\chi_i^* - \chi_i) - \mathbb{E}(\chi_i)
& =
(\hat{\gamma}_i^* - \mathbb{E}(\chi_i) )
\\
& +
\frac{1}{\p} \sum_{\p_1^\prime=1}^{\p} \sum_{\p_2^\prime \neq \p_1^\prime} {X}_{i\varpi_{\p_1^\prime}} {Y}_{i\varpi_{\p_2^\prime}}
\\
& +
\frac{( \p-1 )}{\p} \sum_{\p^\prime=1}^{\p} X_{i\varpi_{\p^\prime}} \, \mathbb{E}^*( \tilde{Y}_{i\varpi} ) 
+
\frac{( \p-1 )}{\p} \sum_{\p^\prime=1}^{\p} Y_{i\varpi_{\p^\prime}} \, \mathbb{E}^*( \tilde{X}_{i\varpi} )
\\
& + 
\left( \bar{\tilde{X}}_i \bar{\tilde{Y}}_i - \mathbb{E}^*(\tilde{X}_{i\varpi}) \, \mathbb{E}^*(\tilde{Y}_{i\varpi}) \right)
+ \T  \left( \mathbb{E}^*(\bar{z}_i^*) \, \mathbb{E}^*(\bar{\varepsilon}_i^*)  -  \bar{z}_i \, \bar{\varepsilon}_i \right),
\end{split}
\end{equation}
where we have used that $\mathbb{E}^*(\tilde{X}_{i\varpi}) = \sqrt{\q} \, \mathbb{E}^*(\bar{z}_i^*)$ and $\mathbb{E}^*(\tilde{Y}_{i\varpi}) = \sqrt{\q} \, \mathbb{E}^*(\bar{\varepsilon}_i^*)$ in the final term on the right-hand side. We now set out to show that, once averaged over $i$, all terms in \eqref{eq:biasde} are $o_{P^*}(1)$.

To deal with the first term in \eqref{eq:biasde} it is helpful to introduce the covariance estimators
$$
\tilde{\gamma}_i
\coloneqq
\frac{1}{\p} \sum_{\p^\prime=1}^{\p} X_{i \varpi_{\p^\prime}} Y_{i \varpi_{\p^\prime}}, \qquad
\hat{\gamma}_i
\coloneqq
\mathbb{E}^*( X_{i\varpi} Y_{i\varpi} ),
$$
and  split the term as
$$
\frac{1}{\N} \sum_{i=1}^\N (\hat{\gamma}_i^* - \mathbb{E}(\chi_i))
=
\frac{1}{\N} \sum_{i=1}^\N ( \hat{\gamma}_i^* - \tilde{\gamma}_i ) 
+
\frac{1}{\N} \sum_{i=1}^\N ( \tilde{\gamma}_i - \hat{\gamma}_i ) 
+
\frac{1}{\N} \sum_{i=1}^\N ( \hat{\gamma}_i - \mathbb{E}( \chi_i ) ).
$$
Here, noting that
$$
\tilde{\gamma}_i - \hat{\gamma}_i^*
=
\left( \frac{1}{\p} \sum_{\p^\prime=1}^{\p} X_{i\varpi_{\p^\prime}} \right)
\left( \frac{1}{\p} \sum_{\p^\prime=1}^{\p} Y_{i\varpi_{\p^\prime}} \right),
$$
an application of the Cauchy-Schwarz inequality, along with Lemma \ref{lemma:Estar} and Lemma \ref{lemma:cross} Part (a) gives
\begin{equation*}
\begin{split}
&\mathbb{E}\left(\mathbb{E}^*\left(\left\lvert \frac{1}{\N} \sum_{i=1}^{\N} ( \hat{\gamma}_i^* - \tilde{\gamma}_i ) \right\rvert \right) \right) 
\\
& \leq
\sup_{i}
\sqrt{
\mathbb{E}\left( \mathbb{E}^*\left( \left\lvert \frac{1}{\p} \sum_{\p^\prime=1}^{\p} X_{i\varpi_{\p^\prime}} \right\rvert^2 \right) \right) 
}
\,
\sqrt{
\mathbb{E}\left( \mathbb{E}^*\left( \left\lvert \frac{1}{\p} \sum_{\p^\prime=1}^{\p} Y_{i\varpi_{\p^\prime}} \right\rvert^2 \right) \right) 
}
\\
& = 
\frac{1}{\p} \sup_{i} 
\sqrt{\mathbb{E}(\mathbb{E}^*(X_{i\varpi}^2))} \, \sqrt{\mathbb{E}(\mathbb{E}^*(Y_{i\varpi}^2))}
\\
& =
O(\q \T^{-1}) . 
\end{split}
\end{equation*}
This deals with the first of the three blocks. For the second block, note that the summands in
$$
\frac{1}{\N} \sum_{i=1}^\N
(\tilde{\gamma}_i - \hat{\gamma}_i)
= 
\frac{1}{\N\p} \sum_{i=1}^\N \sum_{\p^\prime=1}^{\p} ( X_{i\varpi_{\p^\prime}} Y_{i\varpi_{\p^\prime}} - \mathbb{E}^*( X_{i\varpi} Y_{i\varpi} ) )
$$
are mean zero conditional on the data. Hence, by Lemma \ref{lemma:Estar}, 
$$
\mathbb{E}^*((\tilde{\gamma}_i - \hat{\gamma}_i)^2)
=
\frac{1}{\p} \, \mathbb{E}^* (( X_{i\varpi} Y_{i\varpi} - \mathbb{E}^*( X_{i\varpi} Y_{i\varpi} ) )^2)
\leq 
\frac{1}{\p} \, \mathbb{E}^* (X_{i\varpi}^2 Y_{i\varpi}^2) ,
$$
so that, by the Cauchy-Schwarz inequality, 
$$
\mathbb{E}^*((\tilde{\gamma}_i - \hat{\gamma}_i)^2)
\leq \frac{1}{\p} \,  \sqrt{\mathbb{E}^* (X_{i\varpi}^4)} \, \sqrt{\mathbb{E}^* (Y_{i\varpi}^4)}  .
$$
Then, again using the Cauchy-Schwarz inequality, 
$$
\mathbb{E}(\mathbb{E}^*( \lvert \tilde{\gamma}_i - \hat{\gamma}_i \rvert ))
\leq
\frac{1}{\sqrt{\p}} \, \sup_{i}  \sqrt[4]{\mathbb{E}(\mathbb{E}^*(X_{i\varpi}^4))} \,  \sqrt[4]{\mathbb{E}(\mathbb{E}^*(Y_{i\varpi}^4))}
=
O(\sqrt{\q \T^{-1}}), 
$$
by Lemma \ref{lemma:cross} Part (a). For the third and final block, we add and subtract $\mathbb{E}(\hat{\gamma}_i)$ to the summand and use the Cauchy-Schwarz inequality to get 
\begin{equation} \label{eq:inter}
\mathbb{E}\left( \left\lvert \frac{1}{\N} \sum_{i=1}^{\N} ( \hat{\gamma}_i - \mathbb{E}( \chi_i ) ) \right\rvert \right)
\leq
\sqrt{
\mathbb{E}\left( \left( \frac{1}{\N} \sum_{i=1}^{\N} ( \hat{\gamma}_i - \mathbb{E}( \hat{\gamma}_i ) ) \right)^2 \right)
}
+
\sup_{i} \lvert \mathbb{E}( \hat{\gamma}_i ) - \mathbb{E}( \chi_i ) \rvert.
\end{equation}
Now, by cross-sectional independence, 
$$
\mathbb{E}\left( \left( \frac{1}{\N} \sum_{i=1}^{\N} ( \hat{\gamma}_i - \mathbb{E}( \hat{\gamma}_i ) ) \right)^2 \right)
=
\frac{1}{\N^2} \sum_{i=1}^{\N}  \mathbb{E}\left( ( \hat{\gamma}_i - \mathbb{E}( \hat{\gamma}_i ) )^2 \right)
\leq
\frac{1}{\N^2} \sum_{i=1}^{\N}  \mathbb{E}(\hat{\gamma}_i ^2)
\leq
\frac{1}{\N} \sup_{i}  \mathbb{E}(\hat{\gamma}_i ^2).
$$
Here, using the definition of $\hat{\gamma}_i$, the Cauchy-Schwarz inequality, and Lemma \ref{lemma:cross} Part (a) gives
$$
\mathbb{E}( \hat{\gamma}_i^2 )
=
\mathbb{E}\left( \mathbb{E}^*( X_{i\varpi} Y_{i\varpi} )^2 \right)
\leq 
\sqrt{\mathbb{E}( \mathbb{E}^*( X_{i\varpi}^4 ) )} \, \sqrt{\mathbb{E}( \mathbb{E}^*( Y_{i\varpi}^4 ) )} \leq C,
$$
and so the first right-hand side term in \eqref{eq:inter} is $O(\nicefrac{1}{\sqrt{\N}})$. For the second term in this expression, 
$
\mathbb{E}( \hat{\gamma}_i ) = \mathbb{E}( \mathbb{E}^*( X_{i\varpi} Y_{i\varpi} ) )
$
so that  
$
\sup_{i} \lvert \mathbb{E}( \hat{\gamma}_i ) - \mathbb{E}( \chi_i ) \rvert = O( \nicefrac{1}{\q} + \nicefrac{\q}{\T} )
$
by Lemma \ref{lemma:cross} Part (c). Hence, 
$$
	\mathbb{E}\left( \left\lvert \frac{1}{\N} \sum_{i=1}^{\N} ( \hat{\gamma}_i - \mathbb{E}( \chi_i ) ) \right\rvert \right)
	=
	O(\N^{-\nicefrac{1}{2}} + \q^{-1} + \q \T^{-1}) ,
$$
and the analysis of the first term in \eqref{eq:biasde} is complete.

For the second term in \eqref{eq:biasde} we begin by noting that, if $\p_1^\prime \neq \p_2^\prime$ and $\p_3^\prime \neq \p_4^\prime$, we have
$$
\mathbb{E}^*\left( 
X_{i\varpi_{\p_1^\prime}} Y_{i\varpi_{\p_2^\prime}} 
X_{j\varpi_{\p_3^\prime}} Y_{j\varpi_{\p_4^\prime}} 
\right)
=
\left\lbrace
\begin{array}{ll}
\mathbb{E}^*(X_{i\varpi} X_{j\varpi}) \, \mathbb{E}^*(Y_{i\varpi} Y_{j\varpi})  & 
\text{when } \p_1^\prime = \p_3^\prime \text{ and } \p_2^\prime = \p_4^\prime
\\
\mathbb{E}^*(X_{i\varpi} Y_{j\varpi}) \, \mathbb{E}^*(Y_{i\varpi} X_{j\varpi})  &
\text{when } \p_1^\prime = \p_4^\prime \text{ and } \p_2^\prime = \p_3^\prime
\\
0 & \text{otherwise}
\end{array}
\right. .
$$
Hence, 
\begin{equation*}
\begin{split}
\mathbb{E}^*\left( \left( \frac{1}{\N\p} \sum_{i=1}^{\N} \sum_{\p_1^\prime=1}^{\p} \sum_{\p_2^\prime \neq \p_1^\prime} 
X_{i\varpi_{\p_1^\prime}} Y_{i\varpi_{\p_2^\prime}} \right)^2 \right)
& =
\frac{(\p-1)}{\N^2\p}  \sum_{i=1}^{\N} \sum_{j=1}^{\N}
\mathbb{E}^*(X_{i\varpi} X_{j\varpi}) \, \mathbb{E}^*(Y_{i\varpi} Y_{j\varpi}) 
\\
& + 
\frac{(\p-1)}{\N^2 \p}  \sum_{i=1}^{\N} \sum_{j=1}^{\N}
\mathbb{E}^*(X_{i\varpi} Y_{j\varpi}) \, \mathbb{E}^*(Y_{i\varpi} X_{j\varpi}) .
\end{split}
\end{equation*}
The total contribution of summands where $i= j$ is
$$
\frac{( \p-1 )}{\N^2 \p}  \sum_{i=1}^{\N}
\left(
\mathbb{E}^*( X_{i\varpi}^2 ) \, \mathbb{E}^*( Y_{i\varpi}^2 )
+
\mathbb{E}^*( X_{i\varpi} Y_{i\varpi} )^2
\right).
$$
By the Cauchy-Schwarz inequality and Lemma \ref{lemma:cross} the expectation of this sum is bounded from above by 
$$
\frac{( \p-1 )}{\N^2\p}  \sum_{i=1}^{\N}
\left(
\sqrt{\mathbb{E}( \mathbb{E}^*( X_{i\varpi}^2 )^2 )} \, \sqrt{\mathbb{E}( \mathbb{E}^*( Y_{i\varpi}^2 )^2 ) }
+
\mathbb{E}\left( \mathbb{E}^*( X_{i\varpi} Y_{i\varpi} )^2 \right)
\right)  = O(\N^{-1}) .
$$
In the same way, the total contribution of the remaining terms is 
$$
\frac{( \p-1 )}{\N^2 \p}  \sum_{i=1}^{\N} \sum_{j \neq i}
\mathbb{E}\big(
\mathbb{E}^*( X_{i\varpi} X_{j\varpi} ) \, \mathbb{E}^*( Y_{i\varpi} Y_{j\varpi} )
+
\mathbb{E}^*( X_{i\varpi} Y_{j\varpi} ) \, \mathbb{E}^*( Y_{i\varpi} X_{j\varpi} )
\big) 
= 
O(\q \T^{-1}).
$$
Therefore, 
$$
\mathbb{E}\left( \mathbb{E}^*\left( \left\lvert \frac{1}{\N\p} \sum_{i=1}^{\N} \sum_{\p_1^\prime=1}^{\p} \sum_{\p_2^\prime \neq \p_1^\prime} X_{i\varpi_{\p_1^\prime}} Y_{i\varpi_{\p_2^\prime}} \right\rvert \right) \right)
	=
	o( 1 )
$$
follows.

The third term in \eqref{eq:biasde} is mean zero, conditional on the data. Thus 
\begin{equation*}
\begin{split}
\mathbb{E}^*\left( \left( \frac{( \p-1 )}{\N\p} \sum_{i=1}^{\N} \sum_{\p^\prime=1}^{\p} X_{i\varpi_{\p^\prime}} \, \mathbb{E}^*( \tilde{Y}_{i\varpi} ) \right)^2 \right) 
&=
\frac{( \p-1 )^2}{\p\N^2} \sum_{i=1}^{\N} \mathbb{E}^*( X_{i\varpi}^2 ) \, \mathbb{E}^*( \tilde{Y}_{i\varpi} )^2 
\\
&+
\frac{( \p-1 )^2}{\p\N^2} \sum_{i=1}^{\N} \sum_{j \neq i} \mathbb{E}^*( X_{i\varpi} X_{j\varpi} ) \, \mathbb{E}^*( \tilde{Y}_{i\varpi} ) \, \mathbb{E}^*( \tilde{Y}_{j\varpi} ) ,
\end{split}
\end{equation*}
where we have used that $\mathbb{E}^*( X_{i\varpi_{\p_1^\prime}} X_{j\varpi_{\p_2^\prime}} ) = 0$ whenever $\p_1^\prime \neq \p_2^\prime$. 
Now, recalling that $\mathbb{E}^*( \tilde{Y}_{i\varpi} ) = \sqrt{\q} \, \mathbb{E}^*( \bar{\varepsilon}_i^* )$, applying Lemma \ref{lemma:wsum} Part (b) (with $a = 2$ and $w_{it} = \varepsilon_{it}$), we have that 
$$
\sup_{i} \mathbb{E}( \mathbb{E}^*( \tilde{Y}_{i\varpi} )^4 ) = O(\q^2 \T^{-2}). 
$$
Hence, by the Cauchy-Schwarz inequality and Lemma \ref{lemma:cross}, the expectations of (the absolute value of) the two right-hand side terms are bounded by
$$
\frac{( \p-1 )^2}{\p\N^2} \sum_{i=1}^{\N}
\sqrt{\mathbb{E}( \mathbb{E}^*( X_{i\varpi}^2 )^2 )} \,  \sqrt{\mathbb{E}( \mathbb{E}^*( \tilde{Y}_{i\varpi} )^4 ) }
=
O(\N^{-1})
$$
and by
$$
\frac{( \p-1 )^2}{\p\N^2} \sum_{i=1}^{\N} \sum_{j\neq i}
\sqrt{\mathbb{E}( \mathbb{E}^*( X_{i\varpi} X_{j\varpi} )^2 ) } \,
\sqrt[4]{\mathbb{E}( \mathbb{E}^*( \tilde{Y}_{i\varpi} )^4 )} \,
\sqrt[4]{ \mathbb{E}( \mathbb{E}^*( \tilde{Y}_{j\varpi} )^4 )}
=
O(\sqrt{\q\T^{-1}}) , 
$$
respectively.  Hence, 
$$
\mathbb{E}\left( \mathbb{E}^*\left( \left\lvert \frac{( \p-1 )}{\N\p} \sum_{i=1}^{\N} \sum_{\p^\prime=1}^{\p} X_{i\varpi_{\p^\prime}} \, \mathbb{E}^*( \tilde{Y}_{i\varpi} ) \right\rvert \right) \right)
= o( 1 ) 
$$
by the Cauchy-Schwarz inequality. The fourth term in \eqref{eq:biasde} has the same form as the third and can be handled in the same way.

Moving on to the fifth term in \eqref{eq:biasde} we first use the definition of $\tilde{X}_{i\varpi}$ and $\tilde{Y}_{i\varpi}$ to get
$$
\bar{\tilde{X}}_i \bar{\tilde{Y}}_i - \mathbb{E}^*(\tilde{X}_{i\varpi}) \, \mathbb{E}^*(\tilde{Y}_{i\varpi}) 
 =
\q \, \left(\bar{z}_i^* \bar{\varepsilon}_i^* - \mathbb{E}^*(\bar{z}_i^*)  \mathbb{E}^*(\bar{\varepsilon}_i^*) \right).
$$
We then re-arrange this expression as
\begin{equation*}
\begin{split}
\q \, \left(\bar{z}_i^* \bar{\varepsilon}_i^* - \mathbb{E}^*(\bar{z}_i^*)  \mathbb{E}^*(\bar{\varepsilon}_i^*) \right)
& = 
\q \, \left( \bar{z}_i^* - \mathbb{E}^*( \bar{z}_i^* ) \right) \left( \bar{\varepsilon}_i^* - \mathbb{E}^*( \bar{\varepsilon}_i^* ) \right) 
\\
&+ \hspace{.05cm}
\q \, \left( \bar{z}_i^* - \mathbb{E}^*( \bar{z}_i^* ) \right) \, \mathbb{E}^*( \bar{\varepsilon}_i^* )
\\
& +  \hspace{.05cm}
\q \, \left( \bar{\varepsilon}_i^* - \mathbb{E}^*( \bar{\varepsilon}_i^* ) \right) \hspace{.12cm} \mathbb{E}^*( \bar{z}_i^* ).
\end{split}
\end{equation*}
By Lemma \ref{lemma:Estar}, Lemma \ref{lemma:wsum} Part (b), Lemma \ref{lemma:cross} Part (a), and the Cauchy-Schwarz inequality we readily establish that
\begin{equation*}
\begin{split}
\sup_{i} \mathbb{E}( \mathbb{E}^*( \lvert \bar{z}_i^* - \mathbb{E}^*( \bar{z}_i^* ) \rvert^2 ) ) \leq C \, \T^{-1}
& ,
\qquad
\sup_{i} \mathbb{E}( \lvert \mathbb{E}^*( \bar{z}_i^* ) \rvert^2 ) \leq C \, \T^{-1} ,
\\
\sup_{i} \mathbb{E}( \mathbb{E}^*( \lvert \bar{\varepsilon}_i^* - \mathbb{E}^*( \bar{\varepsilon}_i^* ) \rvert^2 ) ) \leq C \, \T^{-1}
& ,
\qquad
\sup_{i} \mathbb{E}( \lvert \mathbb{E}^*( \bar{\varepsilon}_i^* ) \rvert^2 ) \leq C \, \T^{-1} .
\end{split}
\end{equation*}
Therefore, 
$$
\mathbb{E}\left( \mathbb{E}^*\left( \left\lvert \frac{\q}{\N} \sum_{i=1}^{\N} \left( \bar{z}_i^* \bar{\varepsilon}_i^* - \mathbb{E}^*( \bar{z}_i^* ) \, \mathbb{E}^*( \bar{\varepsilon}_i^* ) \right) \right\rvert \right) \right)
= O(\q \T^{-1}) ,
$$
which yields the desired result.

For the sixth and final term in \eqref{eq:biasde} we can proceed in a similar fashion. We first write
$$
\bar{z}_i \bar{\varepsilon}_i - \mathbb{E}^*( \bar{z}_i^* ) \, \mathbb{E}^*( \bar{\varepsilon}_i^* )
=
\left( \bar{z}_i - \mathbb{E}^*( \bar{z}_i^* ) \right) \, \bar{\varepsilon}_i
+
\mathbb{E}^*( \bar{z}_i^* ) \, \left( \bar{\varepsilon}_i - \mathbb{E}^*( \bar{\varepsilon}_i^* ) \right).
$$
Then, initially using the triangle inequality and the Cauchy-Schwarz inequality, followed by an application of Lemma \ref{lemma:wsum} Part (c) (with $w_{it} = z_{it}$ and $w_{it} = \varepsilon_{it}$), Lemma \ref{lemma:wsum} Part (b) (with $w_{it} = z_{it}$ and $a = 1$), and Lemma \ref{lemma:wsum} Part (a) (with $w_{it} = \varepsilon_{it}$, $a = 1$, and $\ell = \T$),
\begin{equation*}
\begin{split}
&
\mathbb{E}\left( \mathbb{E}^*\left( \left\lvert \frac{\T}{\N} \sum_{i=1}^{\N} \left( \bar{z}_i \bar{\varepsilon}_i - \mathbb{E}^*( \bar{z}_i^* ) \, \mathbb{E}^*( \bar{\varepsilon}_i^* ) \right) \right\rvert \right) \right)
\\
& \leq
\T 
\sup_{i} 
\left( \mathbb{E}\left( \lvert \bar{z}_i - \mathbb{E}^*( \bar{z}_i^* ) \rvert \, \lvert \bar{\varepsilon}_i \rvert \right) + \mathbb{E}\left( \lvert \mathbb{E}^*( \bar{z}_i^* ) \rvert \, \lvert \bar{\varepsilon}_i - \mathbb{E}^*( \bar{\varepsilon}_i^* ) \rvert \right) \right)
\\
& \leq
\T \sup_{i} 
\left( 
\sqrt{ \mathbb{E}( \lvert \bar{z}_i - \mathbb{E}^*( \bar{z}_i^* ) \rvert^2 ) \, \mathbb{E}( \bar{\varepsilon}_i^2 ) }
+ 
\sqrt{ \mathbb{E}( \lvert \mathbb{E}^*( \bar{z}_i^* ) \rvert^2 ) \, \mathbb{E}( \lvert \bar{\varepsilon}_i - \mathbb{E}^*( \bar{\varepsilon}_i^* ) \rvert^2 ) 
}
\right)
=
O(\sqrt{\q \T^{-1}}),
\end{split}
\end{equation*}
which yields the desired result.

We have thus shown that each of the terms in \eqref{eq:biasde} is $o_{P^*}( 1 )$, from which it then follows that
$$
- \frac{1}{\N} \sum_{i=1}^{\N} ( \chi_i^* - \chi_i )
=
- \frac{1}{\N} \sum_{i=1}^{\N} \mathbb{E}( \chi_i ) + 	o_{P^*}( 1 ) = 	b_{\N,\T} + o_{P^*}( 1 ),
$$
as claimed.

To complete the proof it remains to show that the first term in \eqref{eq:bexpansion}  follows a central limit theorem. To do so we first let $w_{it} \coloneqq z_{it} \varepsilon_{it}$ and $w_{it}^* \coloneqq z_{it}^* \varepsilon_{it}^*$. Then we are interested in
$$
\frac{1}{\sqrt{\N\T}} \sum_{i=1}^\N \sum_{t=1}^\T (w_{it}^* - w_{it}).
$$
Before proceeding we remark that, by Assumption \ref{ass:ass1} and the definition of $z_{it} = x_{it} - \mathbb{E}(x_{it})$,  
$$
\mathbb{E}(\vert w_{it} \rvert^{2r})^{\nicefrac{1}{2r}} 
\leq
\mathbb{E}(\vert x_{it} \varepsilon_{it} \rvert^{2r})^{\nicefrac{1}{2r}} 
+
\mathbb{E}(\lvert x_{it} \rvert) \, 
\mathbb{E}(\vert \varepsilon_{it}\rvert^{2r})^{\nicefrac{1}{2r}} 
\leq C
$$
for some finite constant $C$ independent of $i$ and $t$. Throughout we will presume that $\q \leq \nicefrac{\T}{2}$.
We begin by writing
$$
\frac{1}{\sqrt{\N\T}} \sum_{i=1}^\N \sum_{t=1}^\T w_{it}^*
=
\frac{1}{\sqrt{\N\T}} \sum_{i=1}^\N \sum_{\p^\prime =1}^{\p} \sum_{\q^\prime =1}^{\q} w_{i\varpi_{\p^\prime}+ \q^\prime}
=
\frac{1}{\sqrt{\p}} \sum_{\p^\prime=1}^\p W_{\varpi_{\p^\prime}},
$$
for
$
W_\varpi \coloneqq \nicefrac{1}{\sqrt{\N\q}} \sum_{i=1}^\N \sum_{\q^\prime =1}^\q w_{i\varpi + \q^\prime}.
$
Then, using that $\T = \p \q$, we observe that we also have
$$
\sqrt{\frac{\T}{\N}} \sum_{i=1}^\N \mathbb{E}^*(\bar{w}_i^*)
=
\sqrt{\frac{\T}{\q}} \, \mathbb{E}^*\left( \frac{1}{\sqrt{\N\q}}  \sum_{i=1}^\N \sum_{\q^\prime=1}^\q w_{i\varpi+\q^\prime} \right)
=
\sqrt{\p} \, \mathbb{E}^*(W_\varpi),
$$
for 
$\bar{w}_i^* \coloneqq \nicefrac{1}{\T} \sum_{t=1}^\T w_{it}^*$ the bootstrap counterpart to 
$\bar{w}_i \coloneqq \nicefrac{1}{\T} \sum_{t=1}^\T  w_{it}$.
This yields the decomposition
$$
\frac{1}{\sqrt{\N\T}} \sum_{i=1}^\N \sum_{t=1}^\T (w_{it}^* - w_{it})
=
\frac{1}{\sqrt{\p}} \sum_{\p^\prime=1}^\p (W_{\varpi_{\p^\prime}} - \mathbb{E}^*(W_\varpi))
+
\sqrt{\frac{\T}{\N}} \sum_{i=1}^\N ( \mathbb{E}^*(\bar{w}_i^*) - \bar{w}_i).
$$
Conditional on the data, the second right-hand side term is non-random. Therefore, by virtue of the triangle inequality,  we can bound 
\begin{equation} \label{eq:sup}
\sup_a 
\left\lvert
\mathbb{P}^*\left( 
\frac{1}{\sqrt{\N\T}} \sum_{i=1}^\N \sum_{t=1}^\T \varOmega_{\N,\T}^{-\nicefrac{1}{2}}  \, (w_{it}^* - w_{it}) \leq a
\right)
- \varPhi(a)
\right\rvert
\end{equation}
from above by 
$$
\sup_a 
\left\lvert
\mathbb{P}^*\left( 
\frac{1}{\sqrt{\p}} \sum_{\p^\prime=1}^\p  {\check{\varOmega}}_{\N,\T}^{-\nicefrac{1}{2}}   \, (W_{\varpi_{\p^\prime}} - \mathbb{E}^*(W_{\varpi})) \leq a
\right)
- \varPhi(a)
\right\rvert 
+
\sup_a 
\left\lvert
\varPhi\left( \frac{a-\check{\mu}_{\N,\T}}{\check{\sigma}_{\N,\T}} \right)
- \varPhi(a)
\right\rvert ,
$$
where $\check{\varOmega}_{\N,\T} \coloneqq  \mathrm{var}^*(W_{\varpi})$, the bootstrap variance of $W_\varpi$, and we introduced the shorthands
$$
\check{\mu}_{\N,\T} \coloneqq 
\sqrt{\frac{\T}{\N}} \sum_{i=1}^\N \varOmega^{-\nicefrac{1}{2}}_{\N,\T} \, (\mathbb{E}^*(\bar{w}_i^*) - \bar{w}_i)
,\qquad 
\check{\sigma}^2_{\N,\T} \coloneqq 
{\varOmega}_{\N,\T}^{-1} \check{\varOmega}_{\N,\T}^{\vphantom{-1}}.
$$
We now show that each of the terms in the upper bound converges to zero in probability.

To show that the first term is $o_P(1)$, let
$$
\check{\varXi}_{\N,\T}  \coloneqq
\mathbb{E}^*(\lvert W_{\varpi} - \mathbb{E}^*(W_\varpi)\rvert^3) .
$$
Because the $W_{\varpi_1},\ldots, W_{\varpi_{\p}}$ are independent and identically distributed conditional on the data, by the Berry-Esseen inequality,  
$$
\sup_a 
\left\lvert
\mathbb{P}^*\left( 
\frac{1}{\sqrt{\p}} \sum_{\p^\prime=1}^\p  {\check{\varOmega}}_{\N,\T}^{-\nicefrac{1}{2}}   \, (W_{\varpi_{\p^\prime}} - \mathbb{E}^*(W_{\varpi})) \leq a
\right)
- \varPhi(a)
\right\rvert 
\leq
C  \, \frac{\check{\varXi}_{\N,\T} \, \check{\varOmega}_{\N,\T}^{-\nicefrac{3}{2}}}{\sqrt{\p}},
$$
whenever $\check{\varOmega}_{\N,\T} >0 $; here $C$ is a constant that does not depend on $\N$, $\T$, or $\p$. We next show that $\check{\varOmega}_{\N,\T} = \varOmega_{\N,\T} + o_{P}(1)$ and that $\check{\varXi}_{\N,\T} = O_P(1)$. Because, by Assumption \ref{ass:ass2}, $\varOmega_{\N,\T}$ is bounded away from zero for all $(\N,\T)$ sufficiently large and $\p = \nicefrac{\T}{\q}$ diverges, the desired result then follows.

We now show that $\check{\varOmega}_{\N,\T} = \varOmega_{\N,\T} + o_{P}(1)$. First, by adding and subtracting $\mathbb{E}(\mathbb{E}^*(W_\varpi^2))$,
$$
\check{\varOmega}_{\N,\T}  -   \varOmega_{\N,\T}
=
\left(
\mathbb{E}(\mathbb{E}^*(W_\varpi^2)) - \varOmega_{\N,\T}
\right)
+
\left(
\mathbb{E}^*(W_\varpi^2)
-
\mathbb{E}(\mathbb{E}^*(W_\varpi^2)) 
\right)
-
\mathbb{E}^*(W_\varpi)^2.
$$
Next, expanding the square and using the definition of $W_\varpi$ allows us to write
$$
W_\varpi^2
=
\frac{1}{\N\q} \sum_{i=1}^\N \left( \sum_{\q^\prime=1}^\q w_{i\varpi+\q^\prime} \right)^2
+
\frac{1}{\N\q} \sum_{i=1}^\N  \sum_{j\neq i}
\sum_{\q_1^\prime=1}^\q  w_{i\varpi+\q_1^\prime}
\sum_{\q_2^\prime=1}^\q  w_{j\varpi+\q_2^\prime},
$$
so that, by Lemma \ref{lemma:Estar},
\begin{equation*}
\begin{split}
\mathbb{E}^*(W^2_\varpi)
& =
\frac{1}{\N\q(\T-\q+1)} \sum_{i=1}^\N \sum_{t=0}^{\T-\q} \left( \sum_{\q^\prime=1}^\q w_{it+\q^\prime} \right)^2
\\
& +
\frac{1}{\N\q(\T-\q+1)} \sum_{i=1}^\N \sum_{j\neq i} \sum_{t=0}^{\T-\q}
 \sum_{\q_1^\prime=1}^\q w_{it+\q_1^\prime}  \sum_{\q_2^\prime=1}^\q w_{jt+\q_2^\prime}.
\end{split}
\end{equation*}
By cross-sectional independence, when $i\neq j$ we have $\mathbb{E}(w_{it+\q_1^\prime} w_{jt+\q_2^\prime}) =0$ and so the second term on the right-hand side is mean zero. Hence,
$$
\mathbb{E}(\mathbb{E}^*(W_\varpi^2))
=
\frac{1}{\N (\T-\q+1)} \sum_{i=1}^\N \sum_{t=0}^{\T-\q} \mathbb{E}\left( \frac{1}{\q} \left(\sum_{\q^\prime=1}^\q w_{it+\q^\prime} \right)^2 \right).
$$
Taken together, the above allow us to decompose the difference $\check{\varOmega}_{\N,\T}  -   \varOmega_{\N,\T}$ as
\begin{equation} \label{eq:clt3}
\begin{split}
\check{\varOmega}_{\N,\T}  -   \varOmega_{\N,\T}
& = 
\mathbb{E}(\mathbb{E}^*(W_\varpi^2)) -  \varOmega_{\N,\T}
\\
& +
\frac{1}{\N\q(\T-\q+1)} \sum_{i=1}^\N \sum_{t=0}^{\T-\q}
\left(
 \left( \sum_{\q^\prime=1}^\q w_{it+\q^\prime} \right)^2
-
\mathbb{E}\left(  \left( \sum_{\q^\prime=1}^\q w_{it+\q^\prime} \right)^2 \right)
\right)
\\
& +
\frac{1}{\N\q(\T-\q+1)} \sum_{i=1}^\N \sum_{j\neq i} \sum_{t=0}^{\T-\q}
 \sum_{\q_1^\prime=1}^\q w_{it+\q_1^\prime}  \sum_{\q_2^\prime=1}^\q w_{jt+\q_2^\prime}
 \\
 & -
 \mathbb{E}^*(W_\varpi)^2 .
\end{split}
\end{equation}
We now show that each of the right-hand side terms vanishes.

For the first term in \eqref{eq:clt3},
$$
\mathbb{E}(\mathbb{E}^*(W_\varpi^2))
=
\frac{1}{\N} \sum_{i=1}^\N
\left(
\mathbb{E}(w_{it}^2) + 2 \sum_{\tau=1}^{\q-1} \left(1 - \frac{\tau}{\q} \right) \, \mathbb{E}(w_{it} w_{it+\tau})
\right)
$$
whereas, recall,
$
\varOmega_{\N,\T}
=
\nicefrac{1}{\N} \sum_{i=1}^\N
(
\mathbb{E}(w_{it}^2) + 2 \sum_{\tau=1}^{\T-1} \left(1 - \nicefrac{\tau}{\T} \right) \, \mathbb{E}(w_{it} w_{it+\tau})
).
$
The difference is, therefore, equal to
$$
\mathbb{E}(\mathbb{E}^*(W_\varpi^2)) -  \varOmega_{\N,\T}
=
-\frac{2}{\N} \sum_{i=1}^\N \sum_{\tau=1}^{\q-1}
\left( \frac{\tau}{\q} - \frac{\tau}{\T} \right) \,  \mathbb{E}(w_{it} w_{it+\tau})
-
\frac{2}{\N} \sum_{i=1}^\N \sum_{\tau=\q}^{\T-1}
\left( 1 - \frac{\tau}{\T} \right) \,  \mathbb{E}(w_{it} w_{it+\tau}).
$$
Again using Davydov's inequality and Assumption \ref{ass:ass1}, 
$$
\sup_{i,t} \sum_{\tau=1}^\infty \tau \, \lvert \mathbb{E}( w_{it} w_{it+\tau} ) \rvert \leq C.
$$
Thus, as $0 \leq \nicefrac{\tau}{\q} - \nicefrac{\tau}{\T} \leq \nicefrac{\tau}{\q}$ for $\tau < \q$ and $0 \leq 1 - \nicefrac{\tau}{\T} \leq 1 \leq \frac{\tau}{\q}$ for $\tau \geq \q$ we have
$$
\left\lvert
\frac{2}{\N} \sum_{i=1}^\N \sum_{\tau=1}^{\q-1}
\left( \frac{\tau}{\q} - \frac{\tau}{\T} \right) \,  \mathbb{E}(w_{it} w_{it+\tau})
\right\rvert
\leq
\frac{2}{\N\q} \sum_{i=1}^\N \sum_{\tau=1}^{\q-1} \tau \, \lvert \mathbb{E}(w_{it} w_{it+\tau}) \rvert
\leq 2 C \q^{-1} ,
$$ 
and
$$
\left\lvert
\frac{2}{\N} \sum_{i=1}^\N \sum_{\tau=\q}^{\T-1}
\left( 1 - \frac{\tau}{\T} \right) \,  \mathbb{E}(w_{it} w_{it+\tau})
\right\rvert
\leq
\frac{2}{\N\q} \sum_{i=1}^\N \sum_{\tau=1}^{\T-1} \tau \, \lvert \mathbb{E}(w_{it} w_{it+\tau}) \rvert
\leq 2 C \q^{-1} ,
$$
from which it follows that 
$$
\left\lvert 
\mathbb{E}(\mathbb{E}^*(W_\varpi^2)) -  \varOmega_{\N,\T}
\right\rvert
=
O(\q^{-1}),
$$
which concludes the analysis for this term.

Moving on, let
$$
A_{it} \coloneqq 
\nicefrac{1}{\q} \left(\sum_{\q^\prime=1}^\q w_{it+\q^\prime} \right)^2
-
\mathbb{E}\left( \frac{1}{\q} \left(\sum_{\q^\prime=1}^\q w_{it+\q^\prime} \right)^2  \right).
$$
Then the second term in \eqref{eq:clt3} can be written as 
$$
\frac{1}{\N(\T-\q+1)} \sum_{i=1}^\N \sum_{t=0}^{\T-\q} A_{it}.
$$
As $A_{it}$ is mean zero and independent over $i$ we have
\begin{equation*}
\begin{split}
\mathbb{E}
\left(\left\lvert
\frac{1}{\N(\T-\q+1)} \sum_{i=1}^\N \sum_{t=0}^{\T-\q} A_{it}
\right\rvert^2 \right)
& =
\frac{1}{\N^2} \sum_{i=1}^\N 
\mathbb{E}
\left(\left\lvert
\frac{1}{(\T-\q+1)}  \sum_{t=0}^{\T-\q} A_{it}
\right\rvert^2 \right)
\\
& \leq
\frac{1}{\N^2(\T-\q+1)} \sum_{i=1}^\N   \sum_{t=0}^{\T-\q} \mathbb{E}(A_{it}^2)
\\
& \leq
\frac{1}{\N\q^2} \sup_{i,t} \mathbb{E}\left( \left(\sum_{\q^\prime=1}^\q w_{it+\q^\prime} \right)^4 \right)
 = O(\N^{-1}),
\end{split}
\end{equation*}
by the Cauchy-Schwarz inequality and an application of Lemma \ref{lemma:wsum} Part (a).

For the third term in \eqref{eq:clt3}, let
$$
B_{ij} \coloneqq \frac{1}{\q(\T-\q+1)} \sum_{t=0}^{\T-\q} \sum_{\q_1^\prime=1}^\q \sum_{\q_2^\prime=1}^\q 
w_{it+\q_1^\prime} w_{jt+\q_2^\prime},
$$
which is symmetric in $(i,j)$. We can then write the relevant term as $\nicefrac{1}{\N} \sum_{i=1}^\N \sum_{j\neq i} B_{ij}$. We have
$$
\mathbb{E}
\left(
\left(
\frac{1}{\N} \sum_{i=1}^\N \sum_{j\neq i} B_{ij}
\right)^2
\right)
=
\frac{1}{\N^2} \sum_{i_1=1}^\N \sum_{j_1\neq i_1}  \sum_{i_2=1}^\N \sum_{j_2\neq i_2}
\mathbb{E}(B_{i_1j_1} \, B_{i_2j_2}).
$$
Here, each $B_{i_1j_1} \, B_{i_2j_2}$ is a sum of products of the form
$w_{i_1t_1} w_{j_1t_2} w_{i_2t_3} w_{j_2 t_4}$. These are mean zero unless $(i_2,j_2)=(i_1,j_1)$ or 
$(i_2,j_2)=(j_1,i_1)$; in either such case, $B_{i_1j_1} = B_{i_2j_2}$, and hence
$$
\mathbb{E}
\left(
\left(
\frac{1}{\N} \sum_{i=1}^\N \sum_{j\neq i} B_{ij}
\right)^2
\right)
=
\frac{2}{\N^2} \sum_{i=1}^\N \sum_{j\neq i} \mathbb{E}(B_{ij}^2)
\lesssim
\sup_{i\neq j} \mathbb{E}(B_{ij}^2).
$$
Now, again using cross-sectional independence, we can obtain the bound 
\begin{equation*}
\begin{split}
\mathbb{E}(B_{ij}^2)
& \leq
\frac{1}{\q^2(\T-\q+1)^2} \sum_{t_1=0}^{\T-\q} \sum_{t_2=0}^{\T-\q}
\left\lvert
\mathbb{E}
\left(
\sum_{\q^\prime_1 =1}^\q
w_{it_1+\q^\prime_1}
\sum_{\q^\prime_2 =1}^\q
w_{it_2+\q^\prime_2}
\right)
\right\rvert
\\
& \times 
\max_{(t_1,t_2)\in \lbrace 0,1,\ldots, \T-\q \rbrace^2}
\left\lvert
\mathbb{E}\left(
\sum_{\q^\prime_1=1}^\q
w_{jt_1+\q^\prime_1}
\sum_{\q^\prime_2=1}^\q
w_{jt_2+\q^\prime_2}
\right)
\right\rvert .
\end{split}
\end{equation*}
We may then proceed exactly as we did in the proof of Lemma \ref{lemma:cross} Part (b) to establish that
$$
\max_{(t_1,t_2)\in \lbrace 0,1,\ldots, \T-\q \rbrace^2}
\left\lvert
\mathbb{E}\left(
\sum_{\q^\prime_1=1}^\q
w_{jt_1+\q^\prime_1}
\sum_{\q^\prime_2=1}^\q
w_{jt_2+\q^\prime_2}
\right)
\right\rvert 
=
O(\q) ,
$$
and
$$
\sum_{t_2=0}^{\T-\q}
\left\lvert
\mathbb{E}
\left(
\sum_{\q^\prime_1 =1}^\q
w_{it_1+\q^\prime_1}
\sum_{\q^\prime_2 =1}^\q
w_{it_2+\q^\prime_2}
\right)
\right\rvert
\leq
\q^2
\,
\sum_{h=-\infty}^\infty \sup_{i,t} \, \lvert \mathbb{E}(w_{it} w_{it+h}) \rvert \leq C \, \q^{2},
$$
using the, by now familiar, argument involving Davydov's inequality applied to $w_{it}$. We thus arrive at
$$
\mathbb{E}
\left(
\left(
\frac{1}{\N} \sum_{i=1}^\N \sum_{j\neq i} B_{ij}
\right)^2
\right)
=
O(\q (\T-\q+1)^{-1}) = O(\q \T^{-1}),
$$
using that $\T < 2(\T-\q+1)$ when $\q < \nicefrac{\T}{2}$.

For the fourth and final term in \eqref{eq:clt3}, because $\mathbb{E}(\mathbb{E}^*(\bar{w}_i^*)) = 0$ and observations are cross-sectionally independent,
$$
\mathbb{E}(\mathbb{E}^*(W_\varpi^*)^2)
=
\frac{\q}{\N} \sum_{i=1}^\N \mathbb{E}(\mathbb{E}^*(\bar{w}_i^*)^2)
= O(\q \T^{-1}),
$$
with the order of magnitude following from an application of Lemma \ref{lemma:wsum} Part (b).

Combining results then yields
$$
\check{\varOmega}_{\N,\T} - {\varOmega}_{\N,\T} = o_{P}(1),
$$
as claimed.

We now show that $\check{\varXi}_{\N,\T} = O_P(1)$. Jensen's inequality and $(x-y)^4 \leq 2^3 (x^4 + y^4)$ give
$$
\check{\varXi}_{\N,\T}
\leq
\left(
\mathbb{E}^*((W_\varpi - \mathbb{E}^*(W_\varpi))^4)
\right)^{\nicefrac{3}{4}}
\leq
2^3 \, \mathbb{E}^*(W_\varpi^4)^{\nicefrac{3}{4}}.
$$
By Lemma \ref{lemma:Estar},
$$
\mathbb{E}(\mathbb{E}^*(W_\varpi^4))
=
\frac{1}{\T-\q+1} \sum_{t=0}^{\T-\q} \mathbb{E}(W_t^4) \leq \sup_{t\in \lbrace 0,1,\ldots, \T-\q \rbrace} \mathbb{E}(W_t^4).
$$
Also, because $\sum_{\q^\prime=1}^\q w_{it+\q^\prime}$ is mean zero and independent in the cross-sectional dimension, 
$$
\mathbb{E}(W_t^4)
=
\frac{1}{\N^2 \q^2} \sum_{i=1}^\N \mathbb{E}\left( \left(\sum_{\q^\prime=1}^\q w_{it+\q^\prime} \right)^4 \right)
+
\frac{3}{\N^2\q^2} \sum_{i=1}^\N \sum_{j\neq i}
\mathbb{E}
\left(
\left(
\sum_{\q^\prime=1}^\q w_{it+\q^\prime}
\right)^2
\,
\left(
\sum_{\q^\prime=1}^\q w_{jt+\q^\prime}
\right)^2
\right)
$$
is bounded uniformly in $t$ by an application of Lemma \ref{lemma:wsum} Part (a) (with $\ell=\q$ and $a=2$ for the first term and $a=1$ for the second term). But then it holds by Markov's inequality that, for any $\epsilon>0$,
$$
\mathbb{P}
\left(
\mathbb{E}^*(W_\varpi^4) > \epsilon
\right)
\leq \frac{\mathbb{E}(\mathbb{E}^*(W_\varpi^4))}{\epsilon} \lesssim \frac{\sup_t \mathbb{E}(W_t^4)}{\epsilon} \leq \frac{C}{\epsilon}.
$$
We can ensure this probability is below any small $\varepsilon >0$ by setting $\epsilon$ slightly larger than $C \varepsilon^{-1}$. Then $\mathbb{E}^*(W_\varpi^4)$ is bounded in probability and
$$
\check{\varXi}_{\N,\T} = O_P(1)
$$
follows. This concludes the proof of the claim that the first term in the upper bound on \eqref{eq:sup} is $o_P(1)$ and it remains only to look at the second term.

To see that
$$
\sup_a 
\left\lvert
\varPhi\left( \frac{a-\check{\mu}_{\N,\T}}{\check{\sigma}_{\N,\T}} \right)
- \varPhi(a)
\right\rvert  = o_P(1)
$$
we begin by noting that
$$
\mathbb{E}
\left(\left\lvert
\sqrt{\frac{\T}{\N}} \sum_{i=1}^\N (\mathbb{E}^*(\bar{w}_i^*) - \bar{w}_i)
\right\rvert^2 \right)
=
\frac{\T}{\N} \sum_{i=1}^\N 
\mathbb{E}(\lvert \mathbb{E}^*(\bar{w}_i^*) - \bar{w}_i \rvert^2)
=
O(\q \T^{-1}) ,
$$
by an application of Lemma \ref{lemma:wsum} Part (c). Then, as $\varOmega_{\N,\T}$ is bounded away from zero for all $(\N,\T)$ sufficiently large, 
$$
\check{\mu}_{\N,\T} =
\varOmega^{-\nicefrac{1}{2}}_{\N,\T} \, 
\left(
\sqrt{\frac{\T}{\N}} \sum_{i=1}^\N (\mathbb{E}^*(\bar{w}_i^*) - \bar{w}_i)
\right)
=
o_P(1).
$$
Furthermore, from above we know that $\check{\sigma}_{\N,\T}^2 = {\varOmega}_{\N,\T}^{-1} \check{\varOmega}_{\N,\T}^{\vphantom{-1}} = 1 + o_P(1)$. Using that $\varPhi$ is a continuous function, the desired result on the second term in the upper bound on \eqref{eq:sup} then follows from P\'olya's theorem. 

Taken together, the last two results yield
$$
\sup_a 
\left\lvert
\mathbb{P}^*\left( 
\frac{1}{\sqrt{\N\T}} \sum_{i=1}^\N \sum_{t=1}^\T \varOmega_{\N,\T}^{-\nicefrac{1}{2}}  \, (w_{it}^* - w_{it}) \leq a
\right)
- \varPhi(a)
\right\rvert
=
o_P(1),
$$
as claimed. This, then, concludes the proof of the theorem.  \qed

\small
\setlength{\bibsep}{0pt} 
\bibliographystyle{chicago3} 
\bibliography{panel}

\end{document}